\documentclass{jfm}
\usepackage[table,dvipsnames]{xcolor}
\usepackage{graphicx}
\usepackage{newtxtext}
\usepackage{newtxmath}
\usepackage[utf8]{inputenc} 
\usepackage[T1]{fontenc} 

\usepackage{natbib}
\usepackage{hyperref}
\usepackage{amsmath}
\usepackage{amssymb}
\usepackage{bm}
\usepackage{cleveref}

\usepackage{framed}
\usepackage{comment}
\hypersetup{
    colorlinks = true,
    urlcolor   = blue,
    citecolor  = black,
}

\newcommand{\RomanNumeralCaps}[1]
\linenumbers

\newcommand{\kB}{k_{\mathrm{B}}}

\newcommand{\Rhc}{\bm{R}_{\mathrm{h}}}
\newcommand{\tRhc}{\tilde{\bm{R}}_{\mathrm{h}}}
\newcommand{\Dhor}{\tilde{D}_{\text{$x$-$y$}}}

\crefname{figure}{figure}{figures}
\crefname{section}{\S\!}{\S\!}
\Crefname{section}{\S\!}{\S\!}
\crefname{appendix}{Appendix}{Appendices}
\Crefname{appendix}{Appendix}{Appendices}

\ifnum 1=1

\makeatletter
\def\ps@titlepage{\leftskip\z@\let\@mkboth\@gobbletwo\vfuzz=5\p@
  \def\@oddhead{\vbox{}} 
  \def\@evenhead{}
  \def\@oddfoot{}
  \def\@evenfoot{}
}
\def\pagelimitfooter{}
\makeatother
\fi 

\title{Taylor Dispersion in Sedimentation of an Axisymmetric Brownian Particle with Centre Offset
}
\author{Zhongqiang Xiong\aff{1},
  Ryohei Seto\aff{1,2,3}
 \and Masao Doi\aff{1,2}\corresp{\email{doi.masao.y3@a.mail.nagoya-u.ac.jp}}}

\affiliation{
\aff{1}Wenzhou Key Laboratory of Biomaterials and Engineering, Wenzhou Institute, University of Chinese Academy of Sciences, Wenzhou 325000, China
\aff{2}Oujiang Laboratory (Zhejiang Lab for Regenerative Medicine, Vision and Brain Health), Wenzhou 325000, China
\aff{3}Graduate School of Information Science, University of Hyogo, Kobe 650-0047, Japan}

\begin{document}

\maketitle

\begin{abstract}

When a non-spherical particle sediments, its velocity generally changes in time as the particle orientation changes in time. This gives extra dispersion of the particle position in addition to the thermal Brownian motion.  Brenner [\textit{J. Colloid Interf. Sci.} 1979, 71(2), 189–208] studied this effect and formulated how to calculate the gravity-induced dispersion (called Taylor dispersion in sedimentation). However, he conducted the explicit calculation only for torque-free particles which keep an isotropic orientational distribution in the steady-state. In this paper, we study the effect of the gravitational torque on the Taylor dispersion. We limit the analysis to particles having uniaxial symmetry. In this case, the gravitational torque is caused by the offset $l_{\mathrm{c}}$, the distance between the hydrodynamic centre and the gravitational force centre. The effect of the gravitational torque is represented by a dimensionless parameter $\alpha$ (called the Langevin parameter by Brenner) which is
proportional to $l_{\mathrm{c}}$. We obtain analytical expressions for the Taylor diffusivity for the two limits, $\alpha \ll 1 $ and $\alpha \gg 1$.  We show that the offset gives a significant effect on the diffusivity and changes the classical scaling of the Taylor dispersion at a large Péclet number.  We also analyze the transient regime of the mean square displacement (MSD) and show how the crossing time from the ballistic regime to the diffusive regime depends on the gravitational torque. 

\end{abstract}


\section{Introduction}

Many realistic particles are non-spherical and experience a non-zero gravitational torque because their centres of mass and buoyancy are offset from the hydrodynamic centre.
These offsets arise naturally in biological systems, such as erythrocytes and plankton, or in particles composed of multiple materials, such as Janus particles.

A three-dimensional position-orientation tracking system has been developed and used to investigate particle sedimentation in experiments. \citet{Roy_Hamati_Tierney_Koch_Voth_2019,PhysRevLett.126.174502,PhysRevFluids.9.070501} performed measurements for non-Brownian particles with asymmetric mass distributions at finite Reynolds numbers.
These studies found that even slight deviations between the centres of mass and buoyancy can dramatically alter settling dynamics. Since translational and rotational motions are generally coupled for non-spherical particles \citep{harvey1980coordinate,Kim_1991,swan2016rapid}, the translation of the particles depends on orientation due to hydrodynamic anisotropy. Therefore, even in the dilute limit, the sedimentation of a non-spherical particle in viscous fluids is a complex phenomenon. 

Sedimentation of a non-spherical Brownian particle in viscous fluids is also complex because the particle rotates and changes its velocity during sedimentation \citep{BRENNER1972228}. This gives extra dispersion of the particle position in addition to the usual Brownian motion. \cite{GOREN1979209} first identified this effect and predicted that the effective translational diffusion depends quadratically on the external force. \cite{brenner1979taylor,brenner1981taylor} later corrected the prefactor in Goren's result and generalized it to uniaxial and triaxial particles, naming it Taylor dispersion in sedimentation. Brenner provided a complete set of equations to compute Taylor dispersion accounting for centre offset, but explicit results have been limited to torque-free particles.
Torque-free particles maintain an isotropic orientation distribution during sedimentation. For such particles, Taylor dispersion increases in proportion to the square of the sedimentation velocity and can become significantly larger than ordinary Brownian dispersion.

In subsequent work, Brenner and colleagues \citep{dill1983general,10.1063/1.450305,pagitsas1986multiple} applied the eigenfunction expansion technique to solve the equations in this general framework, which includes gravitational torques, and to compute the diffusivity for more detailed analysis. Furthermore, Taylor dispersion has been comprehensively reviewed in a classical monograph \citep{brenner1993macrotransport}. Although they demonstrated how to calculate Taylor dispersion with centre offsets, they did not provide quantitative results illustrating how the dispersion depends on the magnitude of the offset or gravitational torque. 
Since Taylor dispersion is the result of the long-time orientation history of particles during sedimentation, the main challenge is that the torque-driven dynamical equations must be solved over long times for different strengths of gravitational torque.

In this paper, we investigate Taylor dispersion for a uniaxial particle subject to gravitational torque in a quiescent Newtonian fluid.
Besides the usage of eigenfunction expansion for numerical calculations, we propose two analytical approaches to solve the dynamical equations at intermediate and large gravitational torques, respectively: an iterative procedure to yield exact solutions for an intermediate torque, and an asymptotic analysis by equating the problem to a Brownian particle trapped in a harmonic potential under large torque. These two methods complement each other and together provide a complete picture. The paper is organized as follows:
In \cref{sec:hydrodynamics}, we analyze the hydrodynamics of an axisymmetric particle sedimenting under gravity taking into account the torque exerted on the hydrodynamic centre by gravity and the buoyancy.
In \cref{sec:Smoluchowski}, we derive the Smoluchowski equation for the distribution function of the particle incorporating these gravitational torque effects.
In \cref{Application}, we apply the Smoluchowski equation to sedimentation-diffusion problems, where orientational dynamics must first be resolved.
In \cref{sec:Rotational}, we derive both the steady-state distribution and transient orientational dynamics.
In \cref{sec:Sedimentation velocity,sec:Horizontal Diffusion}, we analyze the torque-affected sedimentation velocity and Taylor dispersion dynamics, respectively.
In \cref{Transient_MSD}, we compute the transient mean square displacement (MSD) to see reorientation relaxation behavior.

\section{Theoretical Formulation}

\subsection{Hydrodynamics of an Axisymmetric Particle}
\label{sec:hydrodynamics}

\begin{figure}
\centering
\includegraphics[width=0.4\textwidth]{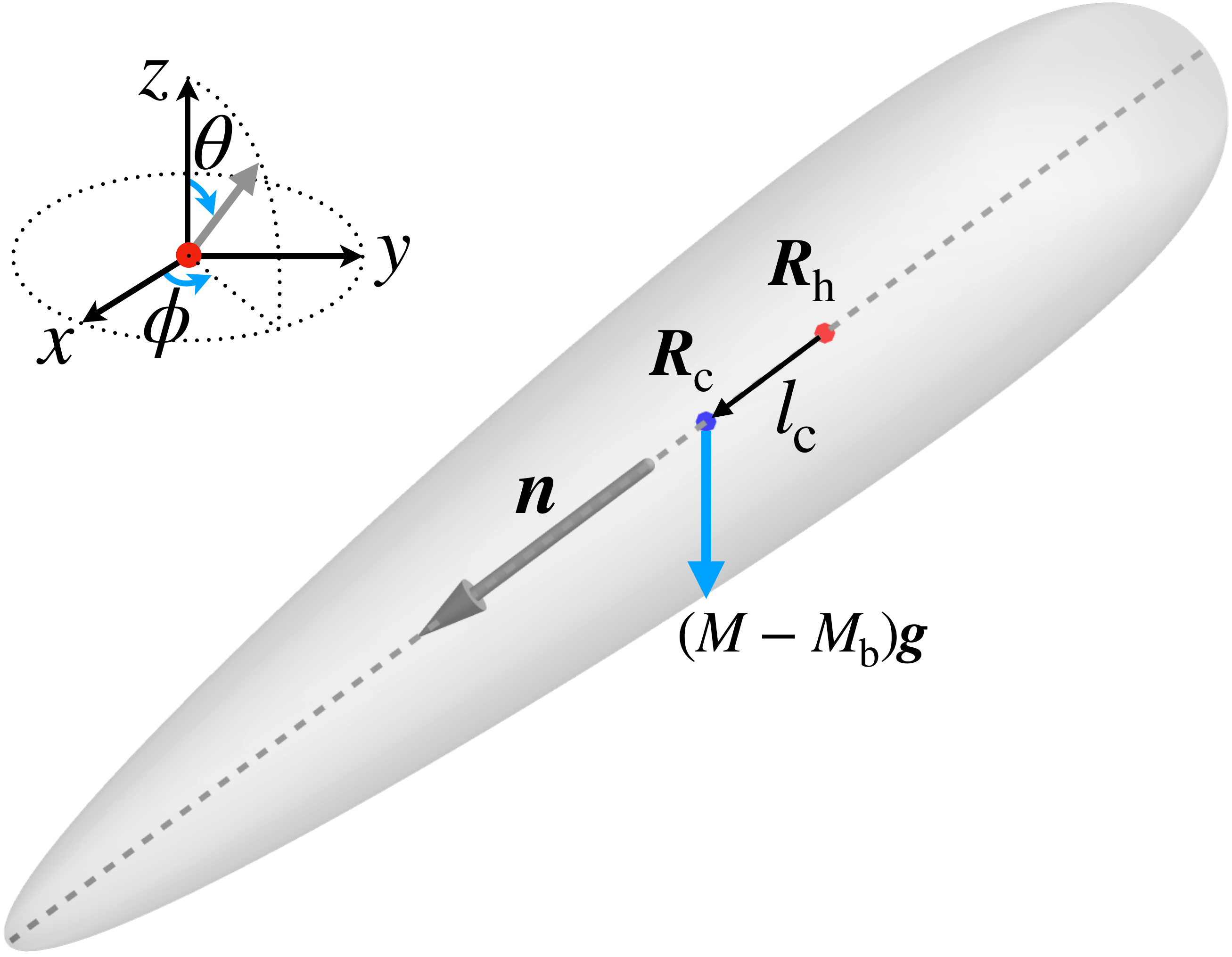}
  \caption{Illustration of an axisymmetric rigid particle, with $\Rhc$ and $\bm{R}_\mathrm{c}$ denoting the hydrodynamic centre and force centre, respectively.
  Vector $\bm{n}$ is the unit vector along the axis of symmetry, and $l_{\mathrm{c}}$ represents the centre offset.
  The force acting at the force centre is the sum of gravity ($M\bm{g}$) and buoyancy ($-M_\mathrm{b}\bm{g}$).}
\label{fig:AB}
\end{figure}

We consider a general axisymmetric rigid particle suspended in a quiescent Newtonian fluid and undergoing Brownian motion, as illustrated in \cref{fig:AB}.
Under Stokes flow conditions ($Re \to 0$), hydrodynamic forces and torques depend linearly on the translational and angular velocities of a particle.
An axisymmetric particle possesses a unique point where translational and rotational motions are decoupled \citep{Kim_1991}: a force applied at this point (regardless of direction) does not produce rotational motion. 
This unique point, known as the hydrodynamic centre $\Rhc$, necessarily lies on the axis of symmetry. 
As long as the surface geometry of the particle is fixed, $\Rhc$ does not vary along the axis and specifies the particle's position. The frictional forces on particles arise from the viscosity of the surrounding fluid.
Referring to $\Rhc$, the hydrodynamic force $\bm{F}_{\mathrm{h}}$ and torque $\bm{T}_{\mathrm{h}}$ acting on an axisymmetric particle can be expressed as
\begin{equation} \label{hydrodynamics}
\begin{bmatrix}
  \bm{F}_{\mathrm{h}}  \\
  \bm{T}_{\mathrm{h}}   
\end{bmatrix}
=-
\begin{bmatrix}
  \bm{\zeta}_\mathrm{t} & \bm{0}  \\
  \bm{0} & \bm{\zeta}_\mathrm{r}  
\end{bmatrix}
\cdot
\begin{bmatrix}
  \bm{u}  \\
  \bm{\omega}  
\end{bmatrix},
\end{equation}
where $\bm{u}$ and $\bm{\omega}$ represent the translational velocity of the hydrodynamic centre and the angular velocity of the particle, respectively.
In terms of time derivatives, we have $\dot{\bm{R}}_{\mathrm{h}}=\bm{u}$ and $\dot{\bm{n}}=\bm{\omega}\times\bm{n}$, where $\bm{n}$ is the unit vector along the symmetry axis.
Because hydrodynamic force depends on particle orientation, 
the translational ($\bm{\zeta}_\mathrm{t}$) and rotational ($\bm{\zeta}_\mathrm{r}$) resistance matrices are expressed through parallel and transverse components relative to the symmetry axis, respectively, i.e.,
\begin{equation}
    \bm{\zeta}_\mathrm{t}=\zeta_\mathrm{t}^\parallel\bm{n}\bm{n}+\zeta_\mathrm{t}^\perp(\bm{\delta}-\bm{n}\bm{n}), \qquad
    \bm{\zeta}_\mathrm{r}=\zeta_\mathrm{r}^\parallel\bm{n}\bm{n}+\zeta_\mathrm{r}^\perp(\bm{\delta}-\bm{n}\bm{n}).
\end{equation}
The four scalar resistance coefficients, $\zeta_\mathrm{t}^\parallel$, $\zeta_\mathrm{t}^\perp$, $\zeta_\mathrm{r}^\parallel$, and $\zeta_\mathrm{r}^\perp$, depend on the particle geometry (i.e., its size and shape) and on the fluid viscosity.
Explicit expressions for the resistance coefficients for prolate and oblate spheroids are tabulated in \cref{resistance}.

Before proceeding, let us consider the properties of gravity and buoyancy for an axisymmetric particle.
Sedimentation arises from density difference between particle and fluid.
Gravitational forces distributed over the particle are mechanically equivalent to a force acting at the centre of mass $\bm{R}_{\mathrm{m}}$, given by 

\begin{equation}
  \bm{R}_{\mathrm{m}} = \frac{1}{M} 
 \int_{V_{\mathrm{p}}}
 d\bm{r} \rho(\bm{r}) \bm{r}, \label{eqn;2.1}
\end{equation}
where $\rho(\bm{r})$ is the local density at position $\bm{r}$ within the particle, 
$M=\int_{V_{\mathrm{p}}} d \bm{r} \rho(\bm{r})$
is the total mass of the particle,
and $ V_{\mathrm{p}}$ denotes the particle volume region.
Similarly, the buoyancy forces are represented by a force acting at the
centre of buoyancy $\bm{R}_{\mathrm{b}}$, given by
\begin{equation}
  \bm{R}_{\mathrm{b}} = 
  \frac{1}{M_{\mathrm{b}}} 
  \int_{V_{\mathrm{p}}} 
  d\bm{r} \rho_0 \bm{r},         \label{eqn;2.20}
\end{equation}
where $\rho_0$ is the density of the surrounding fluid, and $M_{\mathrm{b}}=\rho_0 \int_{V_{\mathrm{p}}} d\bm{r}$ 
is the total mass of the displaced fluid. 
Centres $\bm{R}_{\mathrm{m}}$ and $\bm{R}_{\mathrm{b}}$ must lie on the axis due to the axisymmetry, but they are generally different because they depend on different mass distributions. Nevertheless, the two gravitational forces can be combined further.
The total potential energy of the system is written as
\begin{equation}
   U= 
   -\left(M \bm{g}\cdot \bm{R}_{\mathrm{m}}
   - M_{\mathrm{b}} \bm{g} \cdot \bm{R}_{\mathrm{b}} \right)  = -(M-M_{\mathrm{b}}) \bm{g}\cdot   \bm{R}_{\mathrm{c}}.      \label{eqn;2.3}
\end{equation}
Substituting Eqs.~\eqref{eqn;2.1} and \eqref{eqn;2.20} into Eq.~\eqref{eqn;2.3}, we can combine the effects of gravity and buoyancy acting on the particle, as shown in the second equality, which is equivalently represented by the force $(M-M_{\mathrm{b}}) \bm{g}$ exerted on a point referred to as the \emph{sedimentation force centre} (or simply the \emph{force centre}).
Even though $\bm{R}_{\mathrm{m}}$ and $\bm{R}_{\mathrm{b}}$ are different, the position of the force centre is given by
\begin{equation}
  \bm{R}_{\mathrm{c}} = \frac{1}{M-M_{\mathrm{b}}} 
  \int_{V_{\mathrm{p}}} d\bm{r} [\rho(\bm{r})-\rho_0] \bm{r}.          \label{eqn;2.2}
\end{equation}
The force centre is defined only for particles with $M \neq M_{\mathrm{b}}$. 
In the following, we only consider such particles, since when $M=M_{\mathrm{b}}$, the particle does not sediment and therefore shows no Taylor dispersion.


The explicit expressions for the dispersion coefficients have been obtained for the case where the sedimentation force centre coincides with the hydrodynamic centre, $\bm{R}_{\mathrm{c}} = \Rhc$. 
As illustrated in \cref{fig:AB}, $\bm{R}_\mathrm{c}$ generally differs from $\Rhc$, because $\Rhc$ is exclusively determined by surface geometry, whereas $\bm{R}_\mathrm{c}$ depends on the mass distribution within the particle.
For an axisymmetric particle with $\bm{R}_{\mathrm{c}} \neq \Rhc$, one has
\begin{equation}
	\bm{R}_{\mathrm{c}}=\Rhc+l_{\mathrm{c}}\bm{n}, \label{center_offset}
\end{equation}
since the two centres are on the same axis, where $l_{\mathrm{c}}$ is the centre offset. 
One can also conduct the analyses with the separate offsets of the centres of mass $l_{\mathrm{m}}$ and buoyancy $l_{\mathrm{b}}$ from $\Rhc$ (defined by $\bm{R}_{\mathrm{m}}=\Rhc+l_{\mathrm{m}}\bm{n}$ and $\bm{R}_{\mathrm{b}}=\Rhc+l_{\mathrm{b}}\bm{n}$). The offsets $l_{\mathrm{m}}$ and $l_{\mathrm{b}}$ are all rolled up into $l_{\mathrm{c}}$, which is expressed as 
\begin{equation}
   l_{\mathrm{c}} = \frac{l_{\mathrm{m}} M - l_{\mathrm{b}} M_{\mathrm{b}}}{M - M_{\mathrm{b}}}.     
\end{equation}
Note that $l_{\mathrm{c}}$ is zero even if $l_{\mathrm{m}}$ and $l_{\mathrm{b}}$ are non-zero when $l_{\mathrm{m}} M = l_{\mathrm{b}} M_{\mathrm{b}}$. Adjustments to $l_{\mathrm{c}}$ can be achieved by modifying either the particle's asymmetrical shape or its non-uniform mass distribution. While changes in shape also affect the friction coefficients, changes in mass distribution do not.

\subsection{Smoluchowski Equation under Gravity
}\label{sec:Smoluchowski}

The configuration of the particle is entirely determined by independent variables $\Rhc$ and $\bm{n}$.
Due to the Brownian motion, $\Rhc$ and $\bm{n}$ evolve in a stochastic manner.
Let $\psi( \Rhc,\bm{n},t)$ denote the probability density function of finding the particle at position $\Rhc$ and orientation $\bm{n}$ at time $t$.
The particle dynamics is governed by the hydrodynamic force \eqref{hydrodynamics} and gravitational force (gradient of Eq. \eqref{eqn;2.3}).
Following Onsager's variational principle (\cref{principle}), the probability density $\psi$ evolves according to the Smoluchowski equation:
\begin{multline}
  \frac{\partial \psi}{\partial t}
  =
    D_{\mathrm{r}} \frac{\partial }{\partial \bm{n}}\cdot\left(\bm{\delta}-\bm{n}\bm{n}\right)
    \cdot\left(\frac{\partial \psi}{\partial \bm{n}} - \frac{(M-M_{\mathrm{b}})l_{\mathrm{c}}\bm{g}}{\kB T}\psi\right) \\
   + D^\perp
   \frac{\partial}{\partial \Rhc}\cdot\left(\bm{\delta}
   +
   \frac{\zeta_\mathrm{t}^\perp-\zeta_\mathrm{t}^\parallel}{\zeta_\mathrm{t}^\parallel} \bm{n}\bm{n}\right)\cdot 
    \left(\frac{\partial \psi}{\partial \Rhc}-\frac{(M-M_{\mathrm{b}})\bm{g}}{\kB T}\psi\right),  
    \label{sm_eq}
\end{multline}
where $D_{\mathrm{r}} = \kB T / \zeta_\mathrm{r}^\perp$ is the rotational diffusion coefficient, $D^\perp = \kB T / \zeta_\mathrm{t}^\perp$ is the
transverse diffusion coefficient,
$\kB$ is the Boltzmann constant and $T$ is the temperature.
We adopt the rotational relaxation time $\tau_{\mathrm{r}}$ as the reference time scale:
\begin{equation}
  \tau_{\mathrm{r}} = \frac{1}{D_{\mathrm{r}}} = \frac{\zeta_\mathrm{r}^\perp }{ \kB T}.  
\end{equation}
The rotational relaxation time arises from the rotational diffusion in the absence of gravitational torque. It is the time scale beyond which correlations in the rod orientation become negligible.
In the presence of gravity, the magnitude of the gravitational force and the maximum magnitude of the gravitational torque exerted on particles are given by $F_\mathrm{g}=(M - M_{\mathrm{b}})g$ and $T_\mathrm{g}=(M - M_{\mathrm{b}})gl_{\mathrm{c}}$, respectively, where $g=|\bm{g}|$ denotes the magnitude of the gravitational acceleration. The corresponding characteristic magnitudes of the translational velocities and angular velocity are $u=F_\mathrm{g}/\zeta_{\mathrm{t}}^{\perp}$ and $\omega=T_\mathrm{g}/\zeta_{\mathrm{r}}^{\perp}$, respectively. Therefore, the system has two other characteristic time scales induced by gravitational effects. One is the
\emph{reorientation time} defined by 
\begin{equation}
  \tau_{\mathrm{o}} = \frac{1}{\omega} = 
       \frac{\zeta_{\mathrm{r}}^{\perp}}{ (M-M_{\mathrm{b}})gl_{\mathrm{c}}  },
\end{equation}
which corresponds to the time scale associated with rotation caused by the gravitational torque in the absence of Brownian motion. The other is the \emph{sedimentation time}, defined by
\begin{equation}
 \tau_{\mathrm{s}} = \frac{L}{u} =
\frac{L\zeta_{\mathrm{t}}^{\perp}}{ (M-M_{\mathrm{b}}) g},
\end{equation}
where $L$ is the length scale of the particle, defined by its volume $V$ as $L=[3V/(4\pi)]^{1/3}$. 
It corresponds to the time scale for translation due to the gravitational force in the absence of Brownian motion.
The ratios of rotational relaxation time to gravity-induced time scales, 
$\tau_{\mathrm{r}} / \tau_{\mathrm{o}}$ and  $\tau_{\mathrm{r}} / \tau_{\mathrm{s}}$, 
are two key dimensionless parameters in this system.
They are referred to as the \emph{reorientation Péclet number} $\alpha$ (also referred to as the dimensionless Langevin parameter in \citet{brenner1979taylor}), and the \emph{sedimentation Péclet number} $\beta$, respectively:
\begin{equation}
\alpha = 
\frac{\tau_{\mathrm{r}}}{\tau_{\mathrm{o}}}
=\frac{
(M-M_{\mathrm{b}})gl_{\mathrm{c}}}{
\kB T},
\qquad
\beta = \frac{\tau_{\mathrm{r}}}{\tau_{\mathrm{s}}} 
= 
\frac{\zeta_{\mathrm{r}}^{\perp}}{\zeta_{\mathrm{t}}^{\perp}}
\frac{(M-M_{\mathrm{b}}) g}{L \kB T}. \label{dimensionless_parameter}
\end{equation}
For a very large $\alpha$, the gravitational torque overwhelms rotational Brownian fluctuations, and particle rotation is dominated by gravitational torque rather than rotational diffusion. Similarly, for a very large $\beta$, particle translation is dominated by gravitational force.


We adopt $\tau_{\mathrm{r}}$ and $L$ as the unit of time and length.
Introducing dimensionless time $\tilde{t} = t / \tau_{\mathrm{r}} $ and position 
$ \tRhc = \Rhc / L $, the Smoluchowski equation \eqref{sm_eq} is non-dimensionalised to:
\begin{subequations}\label{Smoluchowski_eqa}
\begin{equation}\label{Smoluchowski_eq}
    \frac{\partial \psi}{\partial \tilde{t}} = \tilde{\mathcal{L}}\psi,
\end{equation}
with
 \begin{equation} \label{oprator}
 	\tilde{\mathcal{L}} = 
      \frac{\partial }{\partial \bm{n}}\cdot\left(\bm{\delta}-\bm{n}\bm{n}\right)\cdot\left(\frac{\partial }{\partial \bm{n}} - \alpha \hat{\bm{g}}\right)
     +   
    \frac{\partial}{\partial \tRhc}\cdot\left(\bm{\delta}+\chi\bm{n}\bm{n}\right)\cdot  \left(\tilde{D}^\perp \frac{\partial}{\partial \tRhc}-\beta \hat{\bm{g}}\right),
 \end{equation}
\end{subequations}
where $\hat{\bm{g}} = \bm{g} /g$, and we have defined
\begin{equation}
  \chi =
    \frac{\zeta_\mathrm{t}^\perp-\zeta_\mathrm{t}^\parallel}{ \zeta_\mathrm{t}^\parallel},
\end{equation}
which characterizes the hydrodynamic anisotropy of the particle.
The dimensionless
transverse diffusion coefficient $\tilde{D}^\perp$ is given by
\begin{equation}
    \tilde{D}^\perp = \frac{D^\perp\tau_{\mathrm{r}}}{L^2} =
\frac{\zeta_\mathrm{r}^\perp}{\zeta_\mathrm{t}^\perp L^2} .\label{DimensionlessDiffusionCoefficient}
\end{equation}
Using the results in Appendix \ref{resistance}, the dimensionless friction-related quantities $\chi$ and $\tilde{D}^\perp$ depend solely on the aspect ratio $r$.


Let $\tilde{\Omega}=(\tRhc,\bm{n})$ denotes the particle configuration at time $\tilde{t}$ and $\tilde{\Omega}'=(\tRhc',\bm{n}')$ denotes the configuration at $\tilde{t}'$. 
Given the initial condition $\left.\psi\right|_{\tilde{t}=\tilde{t}'}=\delta(\tRhc-\tRhc')\delta(\bm{n}-\bm{n}')$, the solution of the Smoluchowski equation \eqref{Smoluchowski_eqa} is the Green's function, which is denoted by $\mathcal{G}(\tilde{\Omega},\tilde{t};\tilde{\Omega}',\tilde{t}')$.
The Green's function represents the conditional probability density of finding the particle in $\tilde{\Omega}$ at time $\tilde{t}$, given that it was in $\tilde{\Omega}'$ at $\tilde{t}'$.
Therefore, the ensemble average of a quantity $\bm{\mathcal{F}}(\tilde{\Omega},\tilde{\Omega}')$ can be generally obtained using the Green's function and initial distribution $\psi_\mathrm{in}$ at $\tilde{t}'$, i.e.,
\begin{equation} \label{EnsembAverage}
 	\langle \bm{\mathcal{F}}(\tilde{t},\tilde{t}')\rangle = \int d\tilde{\Omega} \int d\tilde{\Omega}'\mathcal{G}(\tilde{\Omega},\tilde{t};\tilde{\Omega}',\tilde{t}')\psi_\mathrm{in}(\tilde{\Omega}',\tilde{t}')\bm{\mathcal{F}}(\tilde{\Omega},\tilde{\Omega}'),
\end{equation}
where $d\tilde{\Omega}'=d\tRhc' d\bm{n}'$ is the volume element in the dimensionless configuration space $\tilde{\Omega}'$.


Additionally, one can multiply both sides of the Smoluchowski equation~\eqref{Smoluchowski_eq} by the function $\bm{\mathcal{F}}(\tRhc,\bm{n},\tRhc',\bm{n}')$ and use the definition of the Green's function to obtain the following evolution equation \citep{Doi_1986,macromol4c00532}
\begin{subequations}\label{evlution_eqa}
\begin{equation} \label{evlution_eq}
\frac{\partial \langle \bm{\mathcal{F}}(\tilde{t},\tilde{t}')\rangle}{\partial \tilde{t}} = \langle\tilde{\mathcal{L}}^\dagger \bm{\mathcal{F}}(\tRhc,\bm{n},\tRhc',\bm{n}')\rangle,
\end{equation}
where $\tilde{\mathcal{L}}^\dagger$ is the conjugate operator of $\tilde{\mathcal{L}}$ defined by
\begin{equation}
    \tilde{\mathcal{L}}^\dagger     
= 
 \left(\frac{\partial }{\partial \bm{n}} + \alpha \hat{\bm{g}}\right)\cdot\left(\bm{\delta}-\bm{n}\bm{n}\right)\cdot\frac{\partial }{\partial \bm{n}}
 +
\left(\tilde{D}^\perp\frac{\partial}{\partial \tRhc}
+
\beta\hat{\bm{g}}\right)\cdot
\left(\bm{\delta}+\chi\bm{n}\bm{n}\right)
\cdot\frac{\partial}{\partial \tRhc}.
\end{equation}
\end{subequations}
The ensemble average could also be calculated via Eq.~\eqref{evlution_eqa} without explicitly knowing the Green's function.

\section{Sedimentation and Dispersion}

\subsection{Application of the Smoluchowski Equation}
\label{Application}

We now examine the sedimentation behavior of such a particle in a gravitational field.
Here, we set the gravity direction as $\hat{\bm{g}}=-\bm{e}_z$ and consider the sedimentation process starting from the origin, $\tRhc(0)=\bm{0}$.
The unit basis vectors along the $x$-, $y$-, and $z$-axes are denoted by $\bm{e}_x$, $\bm{e}_y$, and $\bm{e}_z$, respectively.
Since the anisotropy in this system arises solely from gravity, the sedimentation process with horizontal motion can be statistically characterized by the mean position coordinate along the gravity direction $\langle \tilde{z}_{\mathrm{h}}(\tilde{t}) \rangle$, and the mean square displacement (MSD) in the $x$-$y$ plane $\langle \tilde{x}_{\mathrm{h}}^2(\tilde{t}) + \tilde{y}_{\mathrm{h}}^2(\tilde{t})\rangle$.
They are expressed as follows:
\begin{subequations}
\begin{gather}
    \tilde{z}_{\mathrm{h}}(\tilde{t}) =\bm{e}_z\cdot\tRhc(\tilde{t})=
    -\hat{\bm{g}}\cdot[\tRhc(\tilde{t})-\tRhc(0)], 
        \label{Rz}
    \\    \tilde{x}_{\mathrm{h}}^2(\tilde{t})+\tilde{y}_{\mathrm{h}}^2(\tilde{t}) = 
    \left\{(\bm{\delta}-\hat{\bm{g}}\hat{\bm{g}})\cdot[\tRhc(\tilde{t})-\tRhc(0)] \right\}^2.
    \label{RRxy}
\end{gather}
\end{subequations}

The mean position $\langle \tilde{z}_{\mathrm{h}}(\tilde{t})\rangle$ (the drift displacement in the $z$ direction) during sedimentation is obtained by applying Eq.~\eqref{evlution_eq} to Eq.~\eqref{Rz}, yielding
\begin{gather}
	\frac{\partial \langle \tilde{z}_{\mathrm{h}}\rangle
    }{\partial \tilde{t}} 
    = 
    -\left\langle     \tilde{\mathcal{L}}^\dagger\hat{\bm{g}}\cdot[\tRhc(\tilde{t})-\tRhc(0)]
    \right\rangle 
    = 
    -\beta \big(  1+\chi\hat{\bm{g}}\cdot\left\langle \bm{n}\bm{n} \right\rangle\cdot\hat{\bm{g}}
    \big). \label{DRz}
\end{gather}
Similarly, applying Eq.~\eqref{evlution_eq} to the square of Eq.~\eqref{Rz} yields (see \cref{appC} for details)
\begin{align} \label{Rz_sol}
	\frac{\partial \langle \left[\tilde{z}_{\mathrm{h}}-\langle \tilde{z}_{\mathrm{h}}\rangle
	\right]^2\rangle
    }{\partial \tilde{t}} 
    &= 
    2\tilde{D}^\perp \big(  1+\chi\hat{\bm{g}}\cdot\left\langle \bm{n}\bm{n} \right\rangle\cdot\hat{\bm{g}}
    \big) \notag\\
    &\quad +2\left(\beta  \chi\right)^2\int_0^{\tilde{t}}d\tilde{t}'\left[\left\langle \hat{\bm{g}}\cdot\bm{n}\bm{n}\cdot\hat{\bm{g}}\hat{\bm{g}}\cdot\bm{n}'\bm{n}'
		\cdot\hat{\bm{g}}\right\rangle-\hat{\bm{g}}\cdot\left\langle\bm{n}\bm{n}\right\rangle\cdot\hat{\bm{g}}\hat{\bm{g}}\cdot\left\langle \bm{n}'\bm{n}' \right\rangle\cdot\hat{\bm{g}}\right],
\end{align}
where $ \langle \left[\tilde{z}_{\mathrm{h}}-\langle \tilde{z}_{\mathrm{h}}\rangle
\right]^2\rangle$ represents the $z$-direction MSD with eliminating the drift displacement in the $z$ direction.
Here $\bm{n}$ and $\bm{n}'$ denote the orientation of the particle at time $\tilde{t}$ and $\tilde{t}'$, respectively.
The MSD in the $x$-$y$ plane is obtained by applying Eq.~\eqref{evlution_eq} to Eq.~\eqref{RRxy}, yielding
\begin{align} \label{Rxy_sol}
    \frac{\partial \left\langle \tilde{x}_{\mathrm{h}}^2+\tilde{y}_{\mathrm{h}}^2 \right\rangle}{\partial \tilde{t}} 
    &= 
    2\tilde{D}^\perp\left[2+\chi   
    \big(1- \hat{\bm{g}}\cdot\left\langle \bm{n}\bm{n} \right\rangle\cdot\hat{\bm{g}} \big)\right] \notag\\
    &\quad 
    +2\left(\beta  \chi\right)^2\int_0^{\tilde{t}}d\tilde{t}'\left\langle \hat{\bm{g}}\cdot\bm{n}\bm{n}\cdot(\bm{\delta}-\hat{\bm{g}}\hat{\bm{g}})\cdot\bm{n}'\bm{n}'\cdot\hat{\bm{g}}\right\rangle.
\end{align}

\subsection{Orientational Distribution under Gravitational Torque}
\label{sec:Rotational}

\begin{figure}
\centering 
\includegraphics[width=0.5\textwidth]{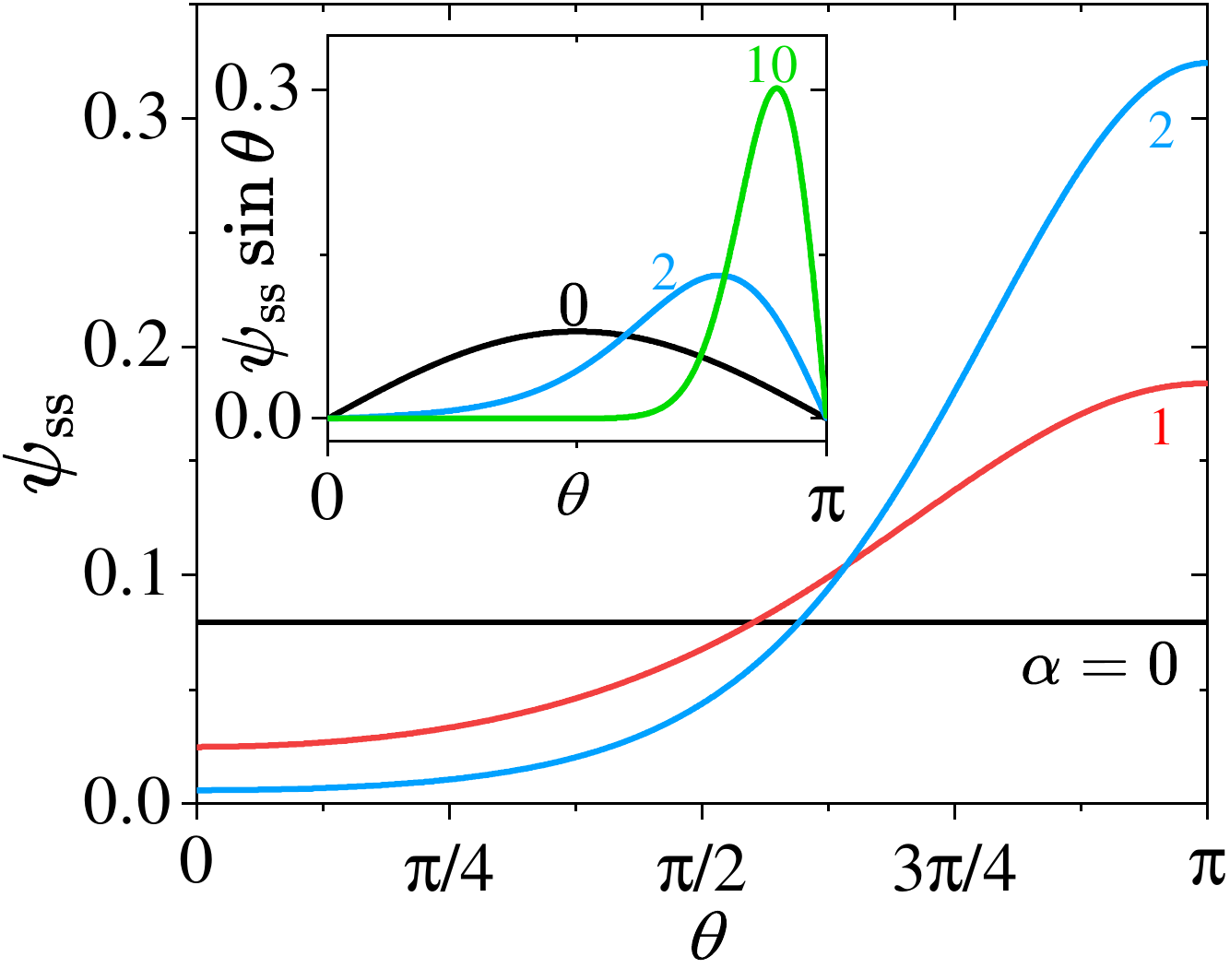}
\caption{Steady-state orientation probability distribution $\psi_{\mathrm{ss}}(\theta, \phi)$ for an axisymmetric Brownian particle.
  Inset: The distribution weighted by $\sin\theta$, accounting for the spherical area element associated with polar angle $\theta$.}
\label{fig:Probability}
\end{figure}

Note that calculation of the sedimentation velocity (from Eq.~\eqref{DRz}) and diffusion (from Eqs.~\eqref{Rz_sol} and \eqref{Rxy_sol}) requires first evaluating the orientation dynamics, especially calculating quantities such as $\left\langle \bm{n}\bm{n} \right\rangle$, $\left\langle \hat{\bm{g}}\cdot\bm{n}\bm{n}\cdot\hat{\bm{g}}\hat{\bm{g}}\cdot\bm{n}'\bm{n}'
		\cdot\hat{\bm{g}}\right\rangle$ and $\left\langle \hat{\bm{g}}\cdot\bm{n}\bm{n}\cdot(\bm{\delta}-\hat{\bm{g}}\hat{\bm{g}})\cdot\bm{n}'\bm{n}'\cdot\hat{\bm{g}}\right\rangle$.
Since these ensemble averages depend solely on orientation, integration over $\tRhc$ can be performed directly.
Defining the probability distribution of the orientation by
\begin{equation}
    \psi(\bm{n},\tilde{t})=\int d\tRhc \psi( \tRhc,\bm{n},\tilde{t}),
\end{equation}
with $d\tRhc = d\tilde{x}_{\mathrm{h}}
d\tilde{y}_{\mathrm{h}}
d\tilde{z}_{\mathrm{h}}$, 
the integration of Eq.~\eqref{EnsembAverage} gives
\begin{equation}
    \langle \bm{\mathcal{F}}(\tilde{t})\rangle = \int d\tilde{\Omega} \psi(\tRhc,\bm{n},\tilde{t})\bm{\mathcal{F}}(\bm{n}) = \int d\bm{n} \psi(\bm{n},\tilde{t})\bm{\mathcal{F}}(\bm{n}).
\end{equation}
The orientation vector $\bm{n}$
is expressed in spherical coordinates as
$\bm{n}=\sin\theta\cos\phi\bm{e}_x
+\sin\theta\sin\phi \bm{e}_y
+\cos\theta \bm{e}_z$ 
with $\theta \in [0,\pi]$ and $\phi \in [0,2\pi]$, illustrated in \cref{fig:AB}.
Then the orientational distribution is expressed as $\psi(\bm{n},\tilde{t})=\psi(\theta,\phi,\tilde{t})$, and $d\bm{n}=\sin\theta d\theta d\phi$.
The equation governing $\psi(\theta,\phi,\tilde{t})$ is obtained by integrating both sides of Eq.~\eqref{Smoluchowski_eq} over $\tRhc$, which reads
\begin{equation}\label{sp}
    \frac{\partial\psi}{\partial\tilde{t}}=\tilde{\mathcal{L}}_\mathrm{sp}\psi = \frac{1}{\sin\theta}\frac{\partial}{\partial\theta}\left(\sin\theta\frac{\partial\psi}{\partial\theta}\right)+\frac{1}{\sin^2\theta}\frac{\partial^2\psi}{\partial\phi^2}-\alpha\left(\sin\theta\frac{\partial\psi}{\partial\theta}+2\cos\theta\psi\right),
\end{equation}
with the periodic boundary condition $\left.\psi\right|_{\phi=0}=\left.\psi\right|_{\phi=2\pi}$ and boundedness conditions of $\psi$ at $\theta=0,\pi$.
The normalization condition is $\int_0^{2\pi}d\phi\int_0^\pi d\theta\sin\theta\psi=1$. 
The steady-state orientation distribution can be obtained analytically by solving $\tilde{\mathcal{L}}_{\mathrm{sp}} \psi_{\mathrm{ss}} = 0$, and the solution is  
\begin{equation}\label{ss}
    \psi_\mathrm{ss}(\theta,\phi) = \frac{\alpha}{4\pi\sinh\alpha}e^{-\alpha\cos\theta},
\end{equation}
(see also \citet{brenner1979taylor} and \citet{dill1983general}).
\Cref{fig:Probability} illustrates the distribution. 
The orientational distribution of a sedimenting particle in a steady-state is the same as the equilibrium distribution of a particle whose hydrodynamic centre is fixed in space.
Due to the rotational symmetry about the $\hat{\bm{g}}$ direction, the steady-state orientation distribution is independent of $\phi$.
When the hydrodynamic centre coincides with the force centre ($l_{\mathrm{c}}=0$), $\alpha$ becomes equal to zero, and Eq.~\eqref{ss} reduces to the uniform distribution $\psi_\mathrm{ss}(\theta,\phi) = 1/(4\pi)$ \citep{frankel1991approach}, which represents a uniform distribution over a unit spherical surface.

Using the steady-state distribution $\psi_\mathrm{ss}$, the original equation~\eqref{sp} can be solved via the Green's function $\mathcal{G}$, determined through an eigenfunction expansion:
\begin{gather}
    \mathcal{G}(\theta,\phi,\tilde{t};\theta',\phi') = \psi_\mathrm{ss}^{-1}(\theta',\phi')\sum_{p=0}^\infty\psi_p(\theta,\phi)\psi_p(\theta',\phi')e^{-\lambda_p\tilde{t}}, \label{Green_n}
\end{gather}
where $\lambda_p$ and $\psi_p$ ($p=0,1,2, \dotsc$) are the eigenvalues and the eigenfunctions of the torque-dependent eigenvalue equation
\begin{equation}
	\tilde{\mathcal{L}}_\mathrm{sp}\psi_p=-\lambda_p\psi_p. \label{eigen_equation}
\end{equation} 
The eigenfunctions satisfy the orthonormality condition $\int_0^{2\pi}d\phi\int_0^\pi d\theta\sin\theta\psi_\mathrm{ss}^{-1}\psi_p\psi_q=\delta_{p,q}$ with weight $\psi_{\mathrm{ss}}^{-1}$.
It can generally be proved that the eigenvalues $\lambda_p$ are all non-negative and real \citep{Doi_1986}.
The sole zero eigenvalue $\lambda_0=0$ is included in Eq.~\eqref{Green_n}, indicating that $\psi(\theta,\phi,\tilde{t})$ ultimately converges to the steady-state distribution $\psi_0=\psi_\mathrm{ss}(\theta,\phi)$ over time.
Other eigenfunctions $\psi_p$ ($p=1,2,3,\dotsc $) are obtained numerically (see \cref{appC}), and they satisfy $\int_0^{2\pi}d\phi\int_0^\pi d\theta\sin\theta\psi_p=0$ for $p=1,2,3,\dotsc $ from the orthonormality.
Therefore, using the Green's function and the initial distribution of orientation $\psi_\mathrm{in}$, the solution of Eq.~\eqref{sp} is
\begin{align}
    \psi(\theta,\phi,\tilde{t}) &= \int_0^{2\pi}d\phi'\int_0^\pi d\theta'\sin\theta'\mathcal{G}(\theta,\phi,\tilde{t};\theta',\phi')\psi_\mathrm{in}(\theta',\phi') \notag\\
    &= \psi_\mathrm{ss}(\theta,\phi) + \sum_{p=1}^\infty c_p \psi_p(\theta,\phi)e^{-\lambda_p\tilde{t}}, \label{solutionPsi}
\end{align}
where $c_p = \int_0^{2\pi}d\phi'\int_0^\pi d\theta'\sin\theta'\psi_\mathrm{ss}^{-1}(\theta',\phi')\psi_\mathrm{in}(\theta',\phi')\psi_p(\theta',\phi')$.

For example, ensemble averages $\langle n_z\rangle$ and $\langle n_z^2\rangle$ are calculated using the solution Eq.~\eqref{solutionPsi}.
The long-time behavior can generally be obtained from the steady-state distribution function \eqref{ss} as
\begin{subequations}
\begin{align}
	\lim_{\tilde{t} \to \infty} \langle n_z(\tilde{t})\rangle &= \langle n_z\rangle_\mathrm{ss} = \int_0^{2\pi}d\phi\int_0^\pi d\theta\sin\theta\psi_\mathrm{ss}(\theta,\phi)\cos\theta=-\frac{\alpha\coth\alpha-1}{\alpha}, \label{lepe} \\
    \lim_{\tilde{t} \to \infty} \langle n_z^2(\tilde{t})\rangle &= \langle n_z^2\rangle_\mathrm{ss} = \int_0^{2\pi}d\phi\int_0^\pi d\theta\sin\theta\psi_\mathrm{ss}(\theta,\phi)\cos^2\theta=1-2\frac{\alpha\coth\alpha-1}{\alpha^2}. \label{cos2}
\end{align}
\end{subequations}
It shows that the particle develops a preferred orientation along the direction of gravity when gravitational torque exists.
Similarly, the steady-state ensemble average $\langle\bm{n}\bm{n}\rangle_\mathrm{ss}$ is obtained by
\begin{equation}
    \langle\bm{n}\bm{n}\rangle_\mathrm{ss} = \frac{1}{3}\bm{\delta}+\left(1-3\frac{\alpha\coth\alpha-1}{\alpha^2}\right)\left(\hat{\bm{g}}\hat{\bm{g}}-\frac{1}{3}\bm{\delta}\right). \label{nnss}
\end{equation}

The transient behavior before the system reaches the steady-state can also be calculated numerically via eigenfunction expansion of the Green's function \eqref{Green_n}, or analytically via perturbation expansion for the moment $\langle(\hat{\bm{g}} \cdot \bm{n})^2\rangle$ with respect to $\alpha$ (see Eq.~\eqref{n}).

For small $\alpha$, the latter method yields the following expression for $\langle n_z(\tilde{t})\rangle$ for a particle which is distributed isotropically at $\tilde t =0$:
\begin{align}\label{n_z_t}
    \langle n_z(\tilde{t})\rangle &= \bm{e}_z\cdot\langle\bm{n}(\tilde{t})\rangle = -\frac{\alpha}{3}
    \big(1-e^{-2\tilde{t}}
    \big)+o(\alpha^2).
\end{align}
Eq.~\eqref{n_z_t} approaches the asymptotic value given by Eq.~\eqref{lepe} as $\tilde{t} \to \infty$.

For large $\alpha$, the gravitational torque overwhelms rotational Brownian fluctuations, locking the particle into a nearly fixed orientation. If the particle is nearly in a  gravity-aligned configuration, one has $n_x,n_y \ll 1$, and $n_z=-[1-(n_x^2+n_y^2)]^{1/2}\approx -1+(n_x^2+n_y^2)/2$. The equilibrium distribution \eqref{ss} is reduced to $\psi_\mathrm{ss} \sim e^{-\alpha n_z} \sim e^{-\alpha(n_x^2+n_y^2)/2} \sim e^{-U_\alpha/(\kB T)}$. This means that the particle experiences a restoring force for a small deviation from the fixed orientation. The restoring force can be evaluated from the gradient of a harmonic potential $U_\alpha=\kB T\alpha(n_x^2+n_y^2)/2$. Therefore, analytical calculation is possible since the dynamics of $\bm{n}(t)$ is the same as the Brownian motion of a particle trapped in a harmonic potential.
The various time correlation functions for $\bm{n}(t)$ are obtained analytically as follows:
\begin{subequations}\label{large_alpha}
\begin{gather}
    \langle n_x(\tilde{t})n_z(\tilde{t})n_x(0)n_z(0)\rangle = \frac{1}{\alpha}e^{-\alpha \tilde{t}}, \label{nx_alpha}\\
    \langle n_z^2(\tilde{t})n_z^2(0)\rangle - \langle n_z^2\rangle^2 = \frac{4}{\alpha^2}e^{-2\alpha \tilde{t}}, \label{nz_alpha}
\end{gather}
\end{subequations}
where the initial equilibrium distribution has been used. These orientational correlation functions are useful for calculating the dispersion coefficients later.

\subsection{Velocity of a Sedimenting Particle}
\label{sec:Sedimentation velocity}

We now turn to the motion along the $z$-direction during sedimentation. This is obtained by using Eq.~\eqref{DRz}.
The sedimentation velocity $\tilde{u}_z$ is defined by
\begin{equation}
    \tilde{u}_z = \lim_{\tilde{t} \to \infty}\frac{\left|\langle \tilde{z}_{\mathrm{h}}(\tilde{t})\rangle \right|}{\tilde{t}}.
\end{equation}
The long-time behavior is given by Eq.~\eqref{DRz}, which then yields
\begin{align}
    \tilde{u}_z &=  
    \beta \big( 1+\chi \hat{\bm{g}}\cdot\langle\bm{n}\bm{n}\rangle_\mathrm{ss}\cdot\hat{\bm{g}} \big) \notag\\
    &=\beta \left[1+\chi\left(1-2\frac{\alpha\coth\alpha-1}{\alpha^2}\right)\right],
\end{align}
where Eq.~\eqref{nnss} for $\langle\bm{n}\bm{n}\rangle_\mathrm{ss}$ has been used.
The sedimentation velocity depends linearly on the sedimentation Péclet number $\beta$.
When $\alpha$ is large, the sedimentation velocity approaches to $\lim_{\alpha\to\infty}\tilde{u}_z = \beta(1+\chi)$, which corresponds to the sedimentation velocity of a rigid particle in a perfectly gravity-aligned configuration.
The dimensional sedimentation velocity is given by $u_z = \tilde{u}_z L/\tau_{\mathrm{r}}$, which reads
\begin{subequations}
\begin{align}
    u_z &= \frac{(M-M_{\mathrm{b}}) g}{\zeta_{\mathrm{t}}^{\perp}}\left[1+\chi\left(1-2\frac{\alpha\coth\alpha-1}{\alpha^2}\right)\right] \\
    &= \frac{(M-M_{\mathrm{b}}) g}{\zeta_{\mathrm{t}}^{\perp}}
    \left[1+\frac{\chi}{3}\left(1+\frac{2\alpha^2}{15}\right)\right]+o(\alpha^2)
    \label{vz_asym}.
\end{align}
\end{subequations}
The second equality \eqref{vz_asym} holds when $\alpha$ is relatively small, indicating that the sedimentation velocity is increasing in $\alpha^2$ for small gravitational torque.

\begin{figure}
\centering
\includegraphics[width=1\textwidth]{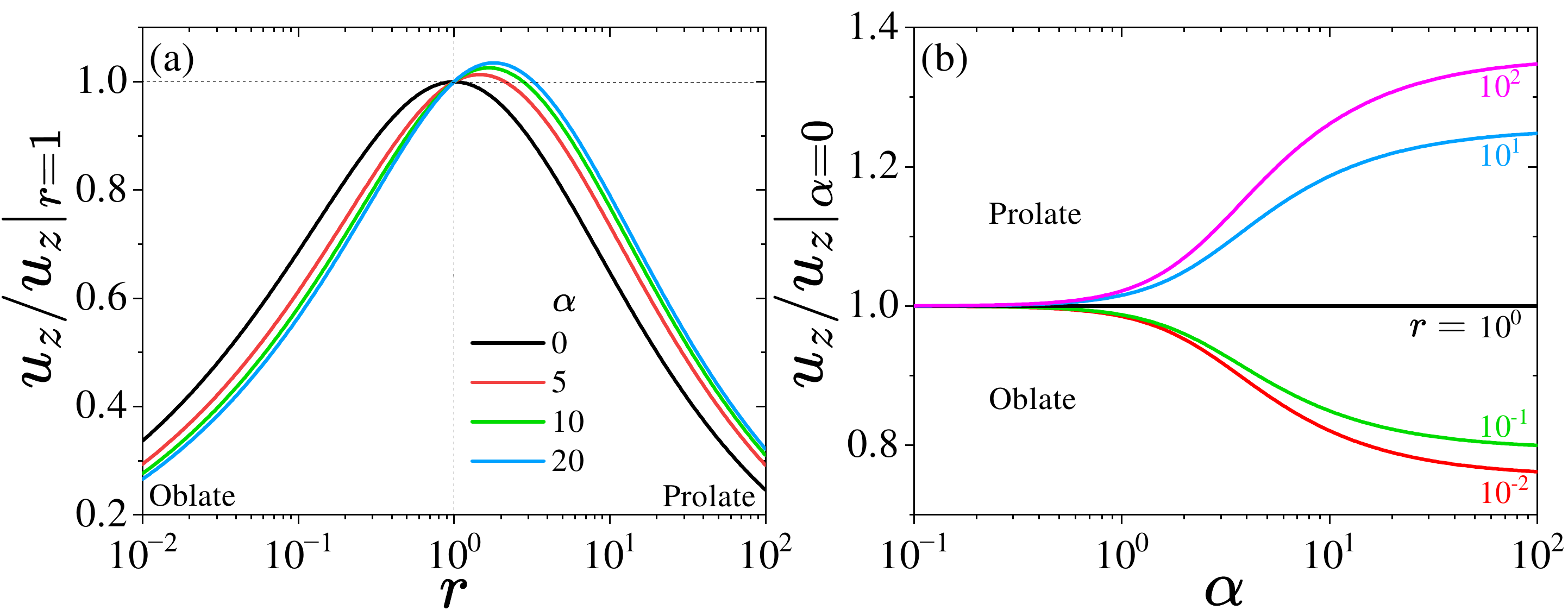}
\caption{Sedimentation velocity:
(a) $u_z$ versus aspect ratio $r$ at various reorientation Péclet numbers $\alpha$;
(b) $u_z$ versus reorientation Péclet number $\alpha$ at various aspect ratios $r$.}
\label{fig:Vz}
\end{figure}

The sedimentation velocity $u_z$ depends on various parameters: particle shape, mass, and gravitational torque $\alpha$. 
As an example, the sedimentation velocities of spheroids (defined by Eq.~\eqref{spheroids}) of constant volume are shown in \cref{fig:Vz} for various aspect ratios $r$ and the gravitational torques $\alpha$.
If the centrosymmetric spheroid density is uniform, the centre offset vanishes (thus gravitational torque $\alpha=0$).
For such particles, it is known that spherical particles exhibit the maximum sedimentation velocity compared with prolate and oblate particles (see the black line in \cref{fig:Vz}\,(a)). 
This is because the average friction constant $[2(\zeta_\mathrm{t}^\perp)^{-1}+(\zeta_\mathrm{t}^\parallel)^{-1}]^{-1}$ of the spheroids is the smallest for a sphere.
However, there is a very narrow band ($1<r\lesssim 4$) where the sedimentation velocity of a prolate spheroid exceeds that of a sphere for non-vanishing $\alpha$.
The reason is that a non-vanishing $\alpha$ leads to an orientational preference along the gravity direction. Since the friction constant $\zeta_\mathrm{t}^\parallel$ is smaller than $\zeta_\mathrm{t}^\perp$ for a prolate spheroid, the sedimentation velocity could be larger than that of a sphere before the average friction constant effect becomes dominant.
\Cref{fig:Vz}\,(b) shows that with the increase of the gravitational torque $\alpha$, the sedimentation velocity increases for prolate but decreases for oblate, which agrees with the early results of \cite{BRENNER1972228}. This difference comes from the friction constant $\zeta_\mathrm{t}^\parallel$ being smaller than $\zeta_\mathrm{t}^\perp$ for prolate but being larger than that for oblate. 

To study the transient sedimentation behavior, we compute the time evolution of $\langle\bm{n}\bm{n}\rangle$.
The ensemble average of $\bm{n}\bm{n}$ can be computed directly from definition \eqref{EnsembAverage} using solution \eqref{solutionPsi} with $\bm{\mathcal{F}} = \bm{n}\bm{n}$.
Alternatively, using Eq.~\eqref{evlution_eq}, the time evolution of the rotational correlation function is expressed as
\begin{align} \label{nn}
    \frac{\partial 
    \langle\bm{n}\bm{n}\rangle
    }{\partial \tilde{t}} 
    = 
    \left\langle
    \tilde{\mathcal{L}}^\dagger\bm{n}\bm{n}
    \right\rangle 
    = 
    -2\left(3\langle\bm{n}\bm{n}\rangle-\bm{\delta}\right)    
    +\alpha\left(
    \hat{\bm{g}}
   \langle\bm{n}\rangle
   +\langle\bm{n}\rangle
    \hat{\bm{g}}-2\hat{\bm{g}}
    \cdot
    \langle\bm{n}\bm{n}\bm{n}
    \rangle\right).
\end{align}

When the hydrodynamic centre and force centre coincide ($\alpha=0$), the exact solution $\langle\bm{n}\bm{n}\rangle = \bm{\delta}/3$ holds for an initially isotropic state.
For $\alpha \neq 0$, we compute the time evolution of $\langle\bm{n}\bm{n}\rangle$ analytically by the perturbation method for small $\alpha$ (details appear in \cref{appA}).
The perturbative expansion gives
\begin{equation} \label{nn_sol}
	\langle\bm{n}(\tilde{t})\bm{n}(\tilde{t})\rangle = \frac{1}{3}\bm{\delta}+\frac{\alpha^2}{15}\left(1-\frac{3}{2}e^{-2\tilde{t}}+\frac{1}{2}e^{-6\tilde{t}}\right)\left(\hat{\bm{g}}\hat{\bm{g}}-\frac{1}{3}\bm{\delta}\right)+o(\alpha^2),
\end{equation}
for a system with initially isotropic distribution. This gives the following perturbation solution:
\begin{align} \label{eq:Rz}
    \langle \tilde{z}_{\mathrm{h}}(\tilde{t})\rangle 
    &= -\tilde{u}_z\tilde{t} 
    + \frac{4\alpha^2}{135} \beta  \chi
      \left(1-\frac{9}{8}e^{-2\tilde{t}}
    + \frac{1}{8}e^{-6\tilde{t}}\right) 
    + o(\alpha^2),
\end{align}
where $\tilde{u}_z$ is now given by Eq.~\eqref{vz_asym}.

\subsection{Dispersion of a Sedimenting Particle}
\label{sec:Horizontal Diffusion}

We now calculate the diffusion in the $x$-$y$ plane and $z$-direction.
As the orientational distribution approaches to the steady-state distribution, the ensemble averages in Eqs.~\eqref{Rxy_sol} and \eqref{Rz_sol} converge to constant values.
We therefore define the horizontal diffusion coefficient $\Dhor$ and the vertical diffusion coefficient $\tilde{D}_z$ by
\begin{subequations}
\begin{gather}
	\Dhor = \lim _{\tilde{t} \to \infty}  \frac{\left\langle \tilde{x}_{\mathrm{h}}^2(\tilde{t})+\tilde{y}_{\mathrm{h}}^2(\tilde{t})\right\rangle}{4 \tilde{t}}, \\
	\tilde{D}_z = \lim _{\tilde{t} \to \infty}  \frac{\langle \left[\tilde{z}_{\mathrm{h}}(\tilde{t})-\langle \tilde{z}_{\mathrm{h}}(\tilde{t})\rangle
\right]^2\rangle}{2 \tilde{t}}.
\end{gather}
\end{subequations}
Equations~\eqref{Rxy_sol} and \eqref{Rz_sol} then yield the following expressions for the horizontal and vertical diffusion coefficients (See Appendix \ref{appC}):
\begin{subequations}
\begin{align}
    \Dhor &=  
    \tilde{D}^\perp\left[1+\frac{\chi}{2}(1-\hat{\bm{g}}\cdot\langle\bm{n}\bm{n}\rangle_\mathrm{ss}\cdot\hat{\bm{g}})\right]
    +\left(\beta  \chi\right)^2\lim_{\tilde{t}\to\infty}\Xi(\tilde{t};\alpha) \notag\\
    &= \tilde{D}^\perp\left(1+\chi\frac{\alpha\coth\alpha-1}{\alpha^2}\right)
    +\left(\beta \chi\right)^2\Xi_\mathrm{ss}(\alpha) , \label{Dxynd}
\end{align}
\begin{align}
    \tilde{D}_z &= \tilde{D}^\perp\left(1+\chi\hat{\bm{g}}\cdot\langle\bm{n}\bm{n}\rangle_\mathrm{ss}\cdot\hat{\bm{g}}\right)
    +\left(\beta  \chi\right)^2\lim_{\tilde{t}\to\infty}\Theta(\tilde{t};\alpha) \notag\\
    &= \tilde{D}^\perp\left(1+\chi-2\chi\frac{\alpha\coth\alpha-1}{\alpha^2}\right)
    +\left(\beta \chi\right)^2\Theta_\mathrm{ss}(\alpha) ,\label{Dznd}
\end{align}
\end{subequations}
where Eq.~\eqref{nnss} for $\langle\bm{n}\bm{n}\rangle_\mathrm{ss}$ has been used. Here $\Xi_\mathrm{ss}(\alpha)$ and $\Theta_\mathrm{ss}(\alpha)$ are defined by
\begin{subequations}
\begin{gather}
    \Xi_\mathrm{ss}(\alpha) = \int_0^{\infty}d\tilde{t} \langle n_x(\tilde{t})n_z(\tilde{t})n_x(0)n_z(0) \rangle, \label{eigenequation} \\
    \Theta_\mathrm{ss}(\alpha) = \int_0^{\infty}d\tilde{t} \left[\langle n_z^2(\tilde{t})n_z^2(0) \rangle-\langle n_z^2 \rangle^2\right], \label{eigenequation2}
\end{gather}
\end{subequations}
which can be calculated numerically by Eqs.~\eqref{Xi_sol} and \eqref{Theta_sol}. 
These are the key quantities to characterize the orientational time correlation in steady-state for the dispersions. However, the analytical expressions of Eq.~\eqref{eigenequation} and \eqref{eigenequation2} are difficult to obtain in general and are only valid for small and large $\alpha$. We can use the iterative method in Appendix \ref{appC} to obtain the expressions in series of $\alpha$. For small $\alpha$, the results up to order of $\alpha^2$ are 
\begin{equation} \label{with_small_alpha}
      \Xi_\mathrm{ss}(\alpha) = \frac{1}{90}\left(1+\frac{59}{252}\alpha^2\right),\qquad
      \Theta_\mathrm{ss}(\alpha) = \frac{2}{135}\left(1+\frac{5}{14}\alpha^2\right) \qquad (\alpha \ll 1).
\end{equation}
Notice that Eq.~\eqref{with_small_alpha} reduces to the classical work of \citet{brenner1979taylor} in the limit of $\alpha \to 0$.

\begin{figure}
\centering 
\includegraphics[width=1\textwidth]{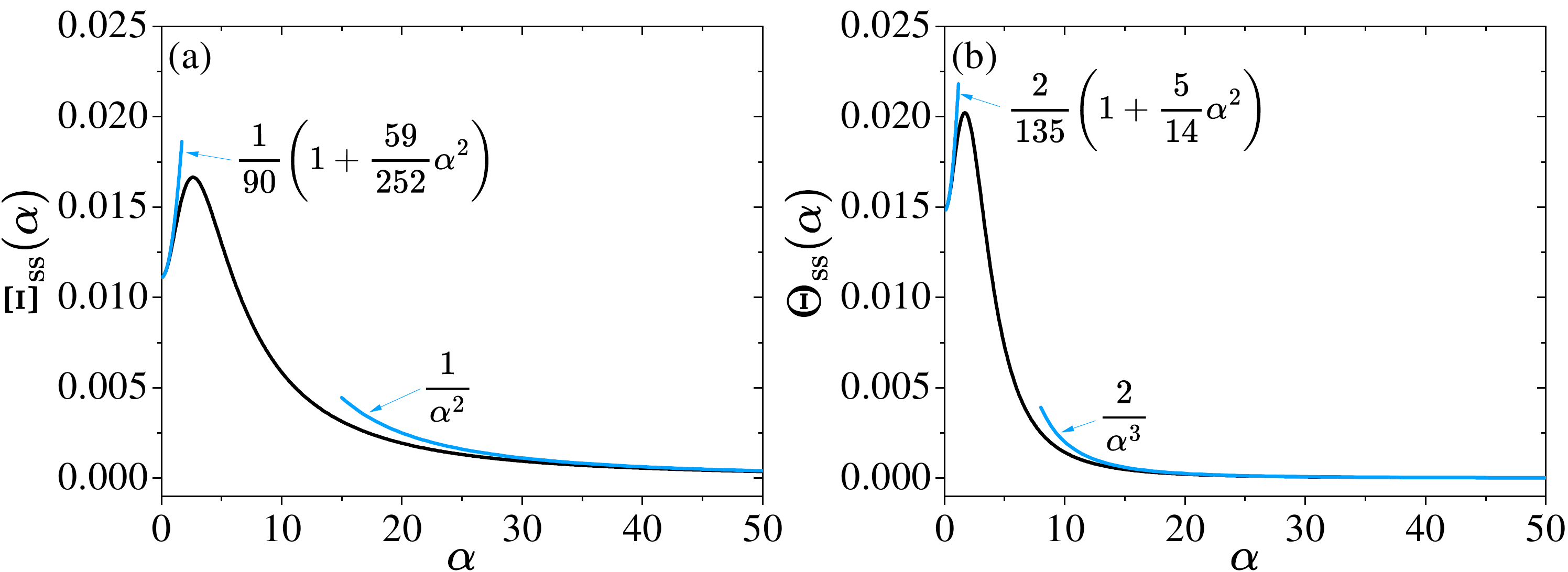}
\caption{Plots of (a) $\Xi_\mathrm{ss}(\alpha)$ defined in Eq.~\eqref{eigenequation} and (b) $\Theta_\mathrm{ss}(\alpha)$ defined in Eq.~\eqref{eigenequation2}. The blue curves show the analytic solutions.}
\label{fig:TD}
\end{figure}

For large $\alpha$, we can use Eq.~\eqref{large_alpha}. The results are 
\begin{equation} 
      \Xi_\mathrm{ss}(\alpha) = \frac{1}{\alpha^2}, \qquad 
      \Theta_\mathrm{ss}(\alpha) = \frac{2}{\alpha^3} \qquad (\alpha \gg 1). \label{XiTheta_la}
\end{equation}
\Cref{fig:TD} illustrates the functions $\Xi_\mathrm{ss}(\alpha)$ and $\Theta_\mathrm{ss}(\alpha)$, where analytic solutions are shown in blue lines.

The apparent diffusivity $\Dhor$ (or $\tilde{D}_z$) originates from two distinct physical mechanisms.
The first mechanism is normal diffusion due to Brownian motion, represented by the first term in Eq.~\eqref{Dxynd} for $\Dhor$ or Eq.~\eqref{Dznd} for $\tilde{D}_z$.
The second mechanism is gravity-induced Taylor dispersion due to hydrodynamic anisotropy, corresponding to the second term in Eq.~\eqref{Dxynd} for $\Dhor$ or Eq.~\eqref{Dznd} for $\tilde{D}_z$.
The torque acting on the particle affects both normal diffusion and Taylor dispersion.
When $\alpha=0$, the well-known Taylor dispersion of \cite{brenner1979taylor} is recovered: the diffusivities have terms proportional to $\beta^2$,  which can be much larger than the terms of Brownian diffusion.
For finite $\beta$, the diffusivities approach limiting values $\lim_{\alpha\to\infty}
\Dhor=\tilde{D}^\perp$ and $\lim_{\alpha\to\infty}
\tilde{D}_z=\tilde{D}^\perp (1+\chi)$. 
These limiting diffusivities correspond to the normal diffusion constants for a rigid particle in a perfectly gravity-aligned configuration.

Note that the original parameter $\beta$ in Eq.~\eqref{dimensionless_parameter} depends on the aspect ratio $r$, and here we define a new parameter  $\beta_0=(M-M_{\mathrm{b}}) gL/(\kB T)$ to represent solely the magnitude of the external force. In addition, the original parameter $\alpha$ in Eq.~\eqref{dimensionless_parameter} depends on the $\beta_0$ as well, and here we define a new parameter $\epsilon=l_\mathrm{c}/L$ to represent solely the dimensionless centre offset. Therefore, we have
\begin{equation}
\alpha=\beta_0\epsilon,
\qquad
\beta=\tilde{D}^\perp\beta_0. \label{dimensionless_parameter_2}
\end{equation}

For completeness, the diffusivities with dimension ($D_{\text{$x$-$y$}} = \Dhor L^2/\tau_{\mathrm{r}}$, $D_z = \tilde{D}_z L^2/\tau_{\mathrm{r}}$; see Eq.~\eqref{DimensionlessDiffusionCoefficient}) are given by
\begin{subequations}
\begin{gather}
    D_{\text{$x$-$y$}} = D^\perp\left(1+\chi\frac{(\beta_0\epsilon)\coth(\beta_0\epsilon)-1}{(\beta_0\epsilon)^2}
    +\tilde{D}^\perp\chi^2\beta_0^2\Xi_\mathrm{ss}(\beta_0\epsilon)\right), \label{Dxy}\\
    D_z = D^\perp\left(1+\chi-2\chi\frac{(\beta_0\epsilon)\coth(\beta_0\epsilon)-1}{(\beta_0\epsilon)^2}
    +\tilde{D}^\perp\chi^2\beta_0^2\Theta_\mathrm{ss}(\beta_0\epsilon)\right), \label{Dz}
\end{gather}
\end{subequations}

\begin{figure}
\centering 
\includegraphics[width=1\textwidth]{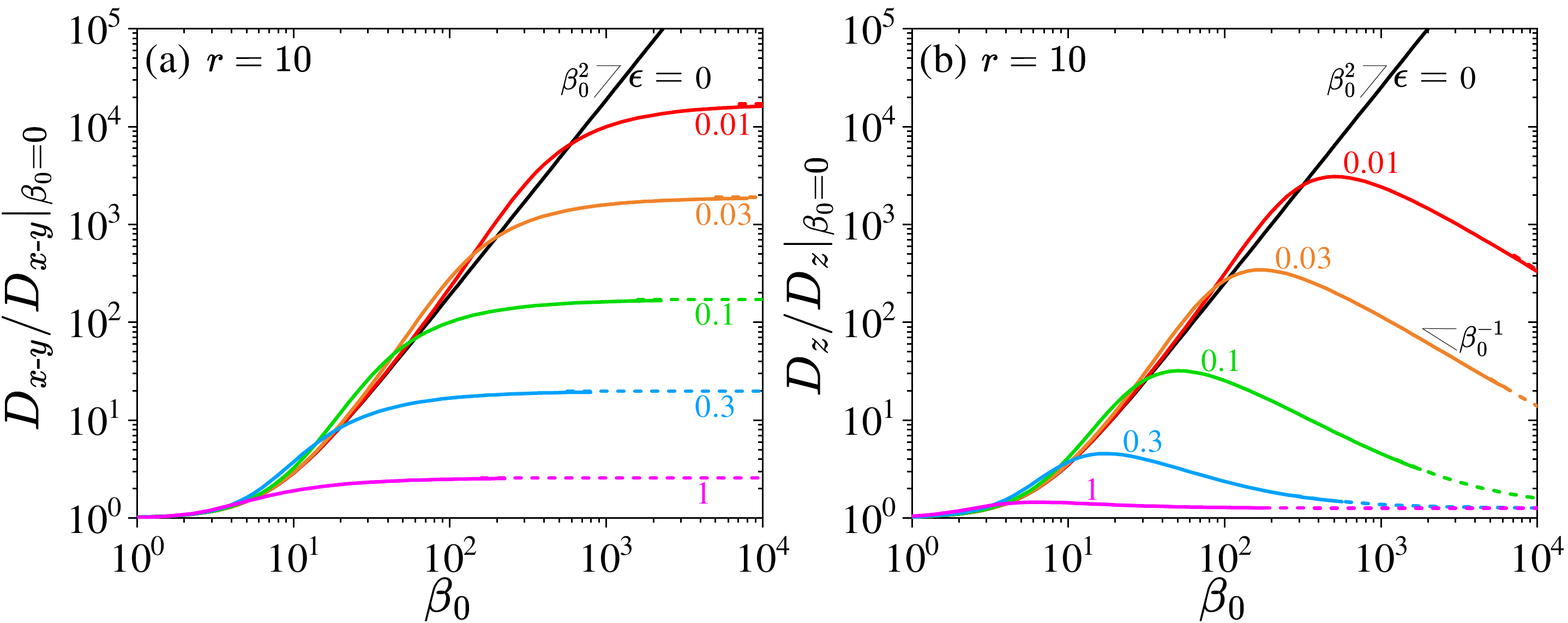}
\includegraphics[width=1\textwidth]{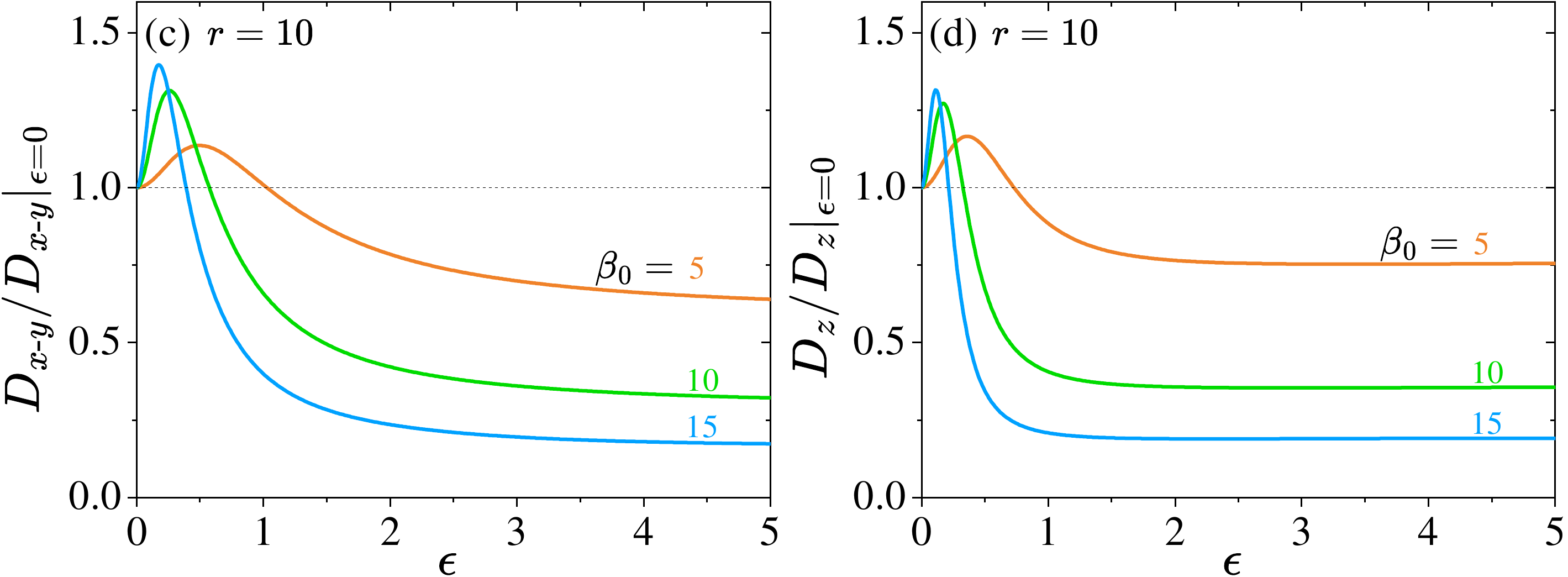}
\caption{(a) The horizontal diffusion coefficient $D_{\text{$x$-$y$}}$ and (b) the vertical diffusion coefficient $D_z$ as functions of the gravity $\beta_0=(M-M_{\mathrm{b}}) gL/(\kB T)$ with fixed dimensionless centre offset $\epsilon=l_\mathrm{c}/L$. The dashed lines are analytical solutions for large $\beta_0$. Coefficients (c) $D_{\text{$x$-$y$}}$ and (d) $D_z$ as functions of $\epsilon$ with fixed $\beta_0$.}
\label{fig:DD}
\end{figure}

For small torque $\beta_0\epsilon \ll 1$, the first-order perturbations of diffusivities $D_{\text{$x$-$y$}}$ and $D_z$ with respect to $(\beta_0\epsilon)^2$ are given by
\begin{subequations}
\begin{gather}
    D_{\text{$x$-$y$}} = D^\perp\left[1+\frac{\chi}{3}\left(1-\frac{(\beta_0\epsilon)^2}{15}\right)
    +\frac{\tilde{D}^\perp\chi^2\beta_0^2}{90}\left(1+\frac{59(\beta_0\epsilon)^2}{252}\right)\right], 
    \label{Dxy_s} \\
    D_z = D^\perp\left[1+\frac{\chi}{3}\left(1+\frac{2(\beta_0\epsilon)^2}{15}\right)
    +\frac{2\tilde{D}^\perp\chi^2\beta_0^2}{135}\left(1+\frac{5(\beta_0\epsilon)^2}{14}\right)\right]. \label{Dz_s}
\end{gather}
\end{subequations}
They indicate that for small torque, the Taylor dispersion of both diffusivities increases with the dimensionless centre offset $\epsilon$.

For large torque $\beta_0\epsilon \gg 1$, the asymptotic behavior of diffusivities $D_{\text{$x$-$y$}}$ and $D_z$ can be calculated using Eq.~\eqref{XiTheta_la}
\begin{subequations}
\begin{gather}
    D_{\text{$x$-$y$}} = D^\perp\left(1+\frac{\chi}{\beta_0\epsilon}
    +\frac{\tilde{D}^\perp\chi^2}{ \epsilon^2}\right), 
    \label{Dxy_l} \\
    D_z = D^\perp\left(1+\chi-\frac{2\chi}{\beta_0\epsilon}
    +\frac{2\tilde{D}^\perp\chi^2}{\beta_0\epsilon^3}\right). \label{Dz_l}
\end{gather}
\end{subequations}
The limit behavior of diffusivities as $\beta_0\to\infty$ with $\epsilon$ fixed can be understood as follows: The diffusivity $D_{\text{$x$-$y$}}$ is estimated by $\langle u_x^2\rangle\tau_\mathrm{o}$ in the ballistic regime of MSD (see \cref{fig:MSDxy} below) since the Péclet number is considerably large, where $u_x$ is the velocity in the $x$ direction. One has $u_x=un_x\sim\beta_0n_x$ since the sedimentation velocity $u$ is proportional to the magnitude of gravity $\beta_0$. Therefore, the diffusivity is given by $D_{\text{$x$-$y$}}\sim \beta_0^2\langle n_x^2\rangle\tau_\mathrm{r}\alpha^{-1} \sim \epsilon^{-2}$, where $\langle n_x^2\rangle \sim \alpha^{-1}$ from the same procedure as in Eq.~\eqref{cos2} and $\alpha=\beta_0\epsilon$ from Eq.~\eqref{dimensionless_parameter_2} have been used. Similarly, the $D_z$ is estimated by $\langle [u_z-\langle u_z\rangle]^2\rangle\tau_\mathrm{o}$ and one has $u_z=un_z\sim\beta_0n_z$. Therefore, the diffusivity is given by $D_z\sim \beta_0^2\langle [n_z-\langle n_z\rangle]^2\rangle\tau_\mathrm{r}\alpha^{-1} \sim \beta_0^{-1}\epsilon^{-3}$, where $\langle [n_z-\langle n_z\rangle]^2\rangle \sim \alpha^{-2}$ from the same procedure as in Eq.~\eqref{cos2} has been used.
  

\begin{figure}
\centering 
\includegraphics[width=0.49\textwidth]{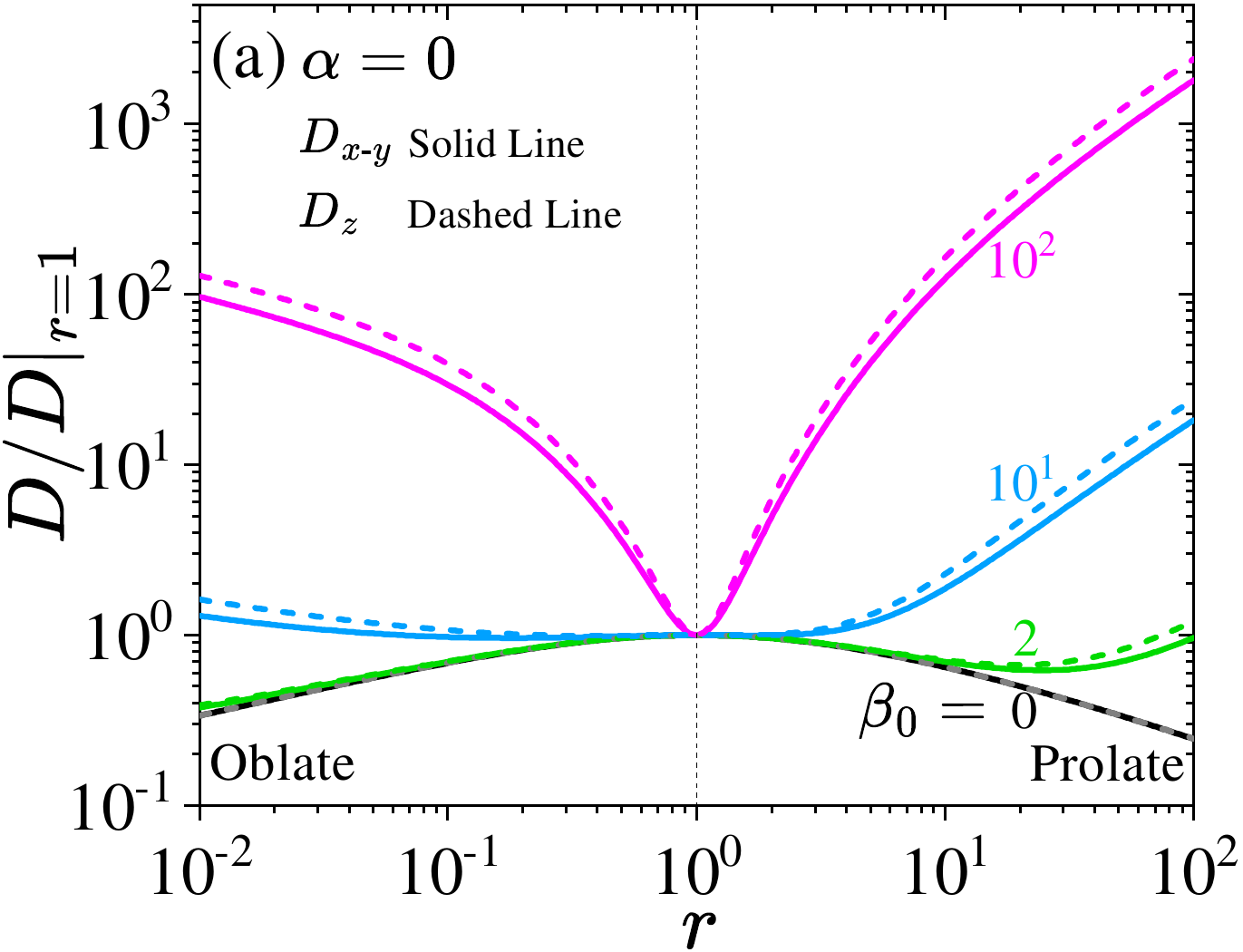}
\includegraphics[width=0.49\textwidth]{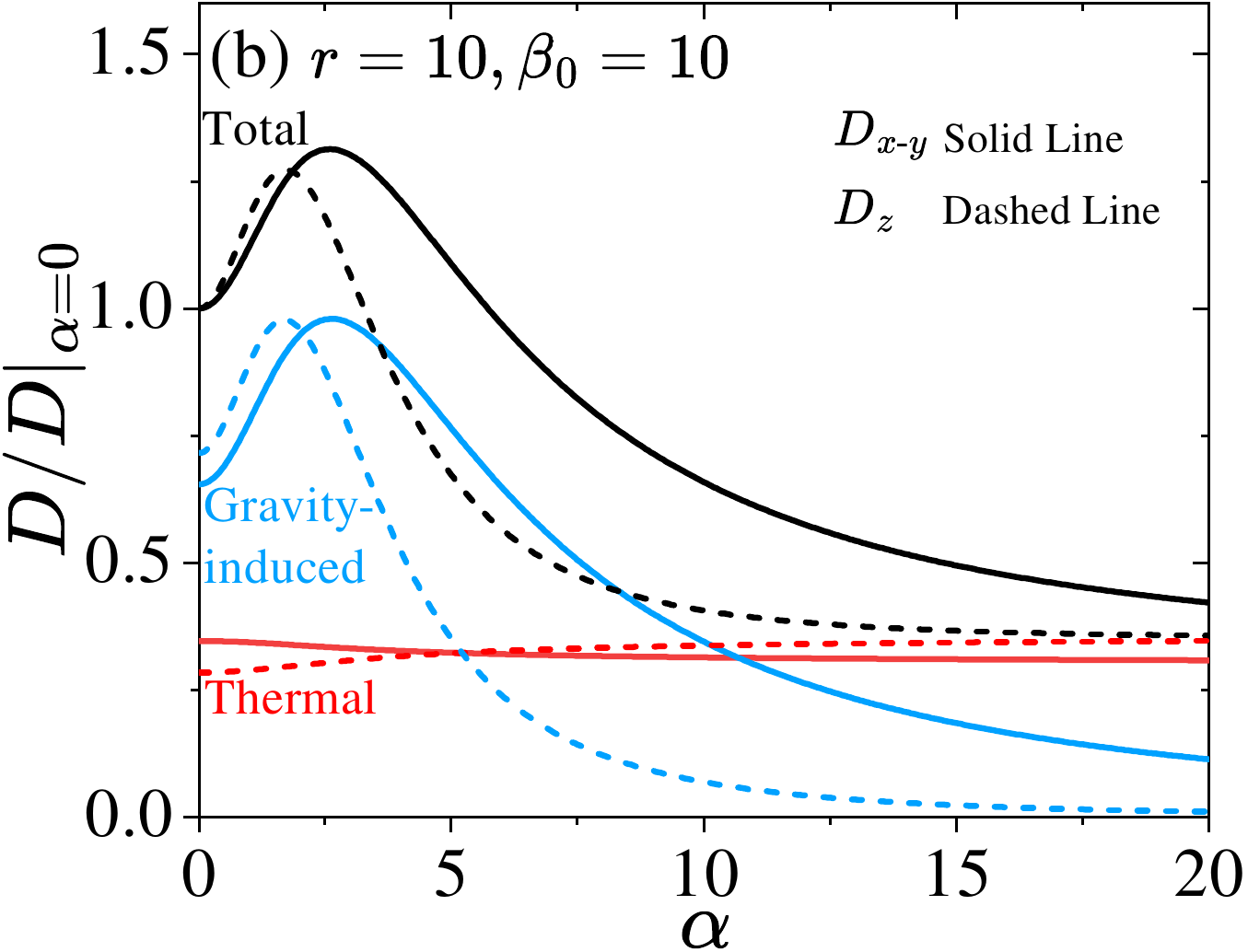}
\includegraphics[width=1\textwidth]{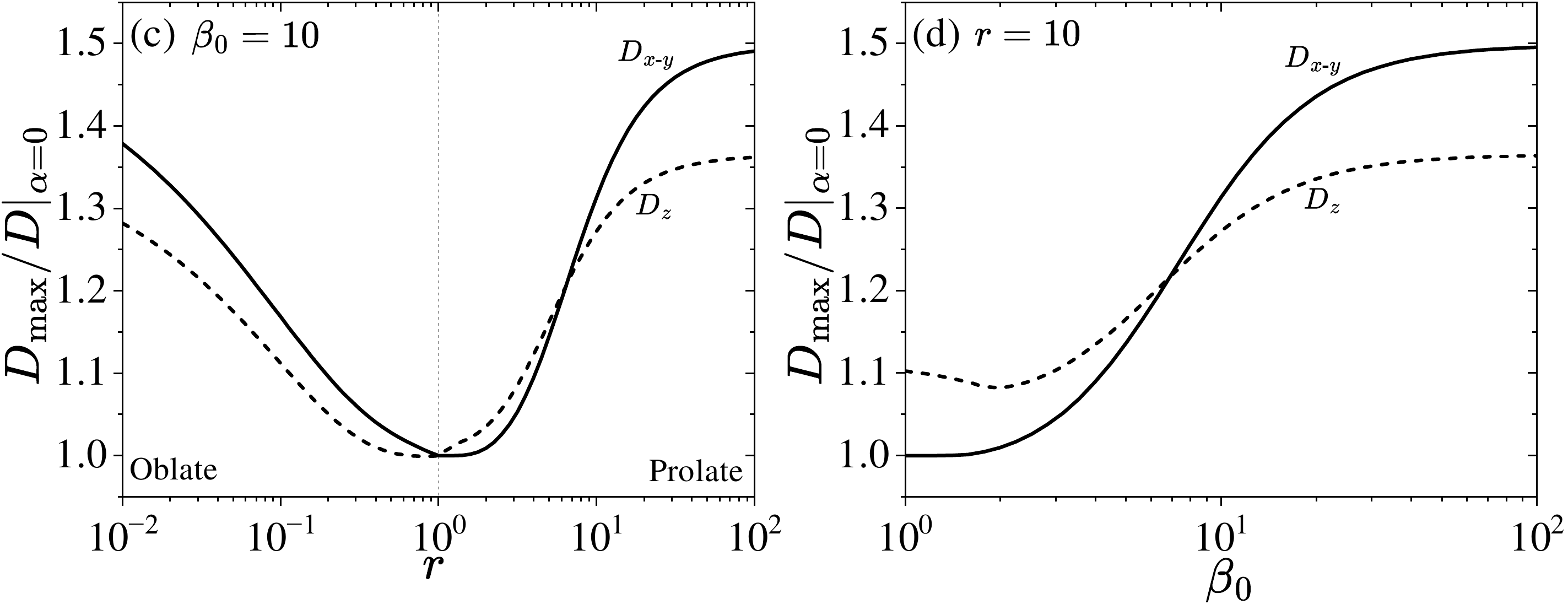}
  \caption{(a) Horizontal ($D_{\text{$x$-$y$}}$) and vertical ($D_z$) diffusion coefficients versus aspect ratio $r$ at $\alpha=0$ (in the absence of gravitational torque). Here, $\beta_0=(M-M_{\mathrm{b}}) gL/(\kB T)$.
  (b) Dependence of horizontal ($D_{\text{$x$-$y$}}$) and vertical ($D_z$) diffusion coefficients on reorientation Péclet number $\alpha$, showing Brownian and gravity-induced contributions.
  The maximum diffusion coefficients $D_\mathrm{max}$ are the peak values of the black lines in panel (b).
  Maximum of horizontal and vertical diffusion coefficients as functions of (c) aspect ratio $r$ and
  (d) sedimentation Péclet number $\beta_0$.}
\label{fig:Dxy}
\end{figure}

\Cref{fig:DD}\,(a) and (b) illustrate, respectively, the horizontal diffusion coefficient $D_{\text{$x$-$y$}}$ in Eq.~\eqref{Dxy} and the vertical diffusion coefficient $D_z$ in Eq.~\eqref{Dz} as functions of the sedimentation Péclet number $\beta_0$ with fixed dimensionless centre offset $\epsilon$. The dashed lines are analytical solutions in Eq.~\eqref{Dxy_l} and Eq.~\eqref{Dz_l} for large $\beta_0$. The black curves show Brenner’s results for the torque-free case. If there is a centre offset with a large $\beta_0$, a significant deviation from the $\beta_0^2$ scaling appears even for a small centre offset.
For sufficiently large $\beta_0$, the horizontal diffusion coefficients $D_{\text{$x$-$y$}}$ approach a constant value which depends on the offset, while the vertical diffusion coefficients $D_z$ all converge to the same value. This indicates that Taylor dispersion is suppressed when the gravitational torque dominates over rotational Brownian fluctuations. 
\Cref{fig:DD}\,(c) and (d) illustrate the $D_{\text{$x$-$y$}}$ and $D_z$ as functions of $\epsilon$ with fixed $\beta_0$, respectively.
The results show that for small centre offset, the Taylor dispersion effect is enhanced compared with the torque-free case, owing to orientational preference induced by the torque. However, if the centre offset is too large, the Taylor dispersion is suppressed since the particle loses the
orientation variability required to generate Taylor dispersion.

\Cref{fig:Dxy}\,(a) shows the results of \cite{brenner1979taylor} for the
situation of $\alpha=0$, i.e., the force centre is at the hydrodynamic centre. A similar graph was shown by \cite{GOREN1979209}.
It is known that for particles of constant volume, the diffusivity $D_{\text{$x$-$y$}}$ (or $D_z$) of neutrally buoyant ($\beta_0=0$) particles is maximized for spherical particles, since the average friction constant $[2(\zeta_\mathrm{t}^\perp)^{-1}+(\zeta_\mathrm{t}^\parallel)^{-1}]^{-1}$ is smallest for spherical particles.
For large $\beta_0$, gravity-induced Taylor dispersion dominates over ordinary diffusion, but no Taylor dispersion occurs for spheres. Since the coefficient $D^{\perp}/(D_{\mathrm{r}}L^2)$, scales as $r^{4/3}$ for $r \gg 1$ and $r^{-2/3}$ for $r \ll 1$, the diffusivity is minimized for spheres.  
Therefore, for sedimenting particles with intermediate $\beta_0$, the competition appears between the ordinary diffusion and the Taylor dispersion so that the minimum occurs slightly over into the prolate range.
Both $D_{\text{$x$-$y$}}$ and $D_z$ exhibit similar behavior.

In \cref{fig:Dxy}\,(b), the diffusivities of particles with gravitational torque ($\alpha \neq 0$)
normalized by that without torque are shown as a function of $\alpha$, where aspect ratio $r=10$ and sedimentation Péclet number $\beta_0=10$ are chosen.
Here contributions of the two mechanisms, the normal diffusion caused by thermal motion
and the Taylor dispersion induced by gravity, are shown separately.
It is seen that for small gravitational torque, the Taylor dispersion is enhanced, but 
as $\alpha$ increases, the Taylor dispersion starts to be suppressed. 
This occurs because the gravitational torque overwhelms rotational Brownian fluctuations, locking the particle into a nearly fixed orientation and thereby removing the orientational variability required to generate Taylor dispersion.
As a consequence both 
$D_{\text{$x$-$y$}}$ and $D_z$ show maximum at a non-zero value of $\alpha$.
As $\alpha$ goes to infinity, the diffusivities approach to their asymptotic values, 
i.e., $D_{\text{$x$-$y$}} \to  D^\perp$ and $D_{\text{$z$}} \to D^\perp(1+\chi)$. 


The $\alpha$ dependence of the diffusivities changes when the other parameters changes. 
The maximum diffusion coefficients $D_\mathrm{max}$ are defined as the peak values of the black lines in \cref{fig:Dxy}\,(b).
The maximum diffusivity takes place around the same value of $\alpha \sim 2$ for both prolate and oblate spheroids, as well as for both horizontal and vertical diffusivities. 
Note that the maximum diffusivity at non-vanishing $\alpha$ exceeds the classical Taylor dispersion (at $\alpha=0$) due to the orientational preference of the particles.
The maximum diffusivity itself varies with aspect ratio $r$ and sedimentation Péclet number $\beta_0$.
\Cref{fig:Dxy}\,(c) and (d) show the maximum diffusivity as a function of $r$ and of $\beta_0$, respectively. The enhanced Taylor dispersion is larger than the Taylor dispersion without a centre offset and saturates at approximately $150\%$ of the dispersion without centre-offset. This implies that one can further enhance particle dispersion by tuning $\alpha$ to $\sim 2$, for example, by adjusting the mass distribution of the particles.

\subsection{Transient Behavior of Mean Square Displacement} \label{Transient_MSD}

We now compute the transient MSD to characterize dynamic evolution.
First we consider the case of $\alpha=0$. 
Since the orientational distribution is isotropic when $\alpha=0$, the right-hand sides of Eqs.~\eqref{Rxy_sol} and \eqref{Rz_sol} can be computed exactly for a system with initially isotropic distribution.
The results of the horizontal and vertical MSD are
\begin{subequations}\label{hMSD_exact_all}
\begin{gather}
	\langle \tilde{x}_{\mathrm{h}}^2(\tilde{t})+\tilde{y}_{\mathrm{h}}^2(\tilde{t}) \rangle = 4\left[\tilde{D}^\perp\left(1+\frac{\chi}{3}\right)
    +\frac{\left(\beta \chi\right)^2}{90}\right] \tilde{t}-
    \frac{1}{15} \left(\frac{\beta  \chi}{3}\right)^2\left(1-e^{-6\tilde{t}}\right),
    \label{hMSD_exact} \\
    \langle \left[\tilde{z}_{\mathrm{h}}(\tilde{t})-\langle \tilde{z}_{\mathrm{h}}(\tilde{t})\rangle
	\right]^2\rangle = 2\left[\tilde{D}^\perp\left(1+\frac{\chi}{3}\right)
    +\frac{2\left(\beta \chi\right)^2}{135}
    \right] \tilde{t}-
    \frac{2}{45} \left(\frac{\beta  \chi}{3}\right)^2\left(1-e^{-6\tilde{t}}\right).
\end{gather}
\end{subequations}

The transient MSD for $\alpha \neq 0$ is obtained from Eq.~\eqref{Rxy} (or \eqref{Rzz}) by perturbation up to $\alpha^2$ using Laplace transform methods (Eqs.~\eqref{TFH}).


\begin{figure}
  \centering
  \includegraphics[width=0.48\textwidth]{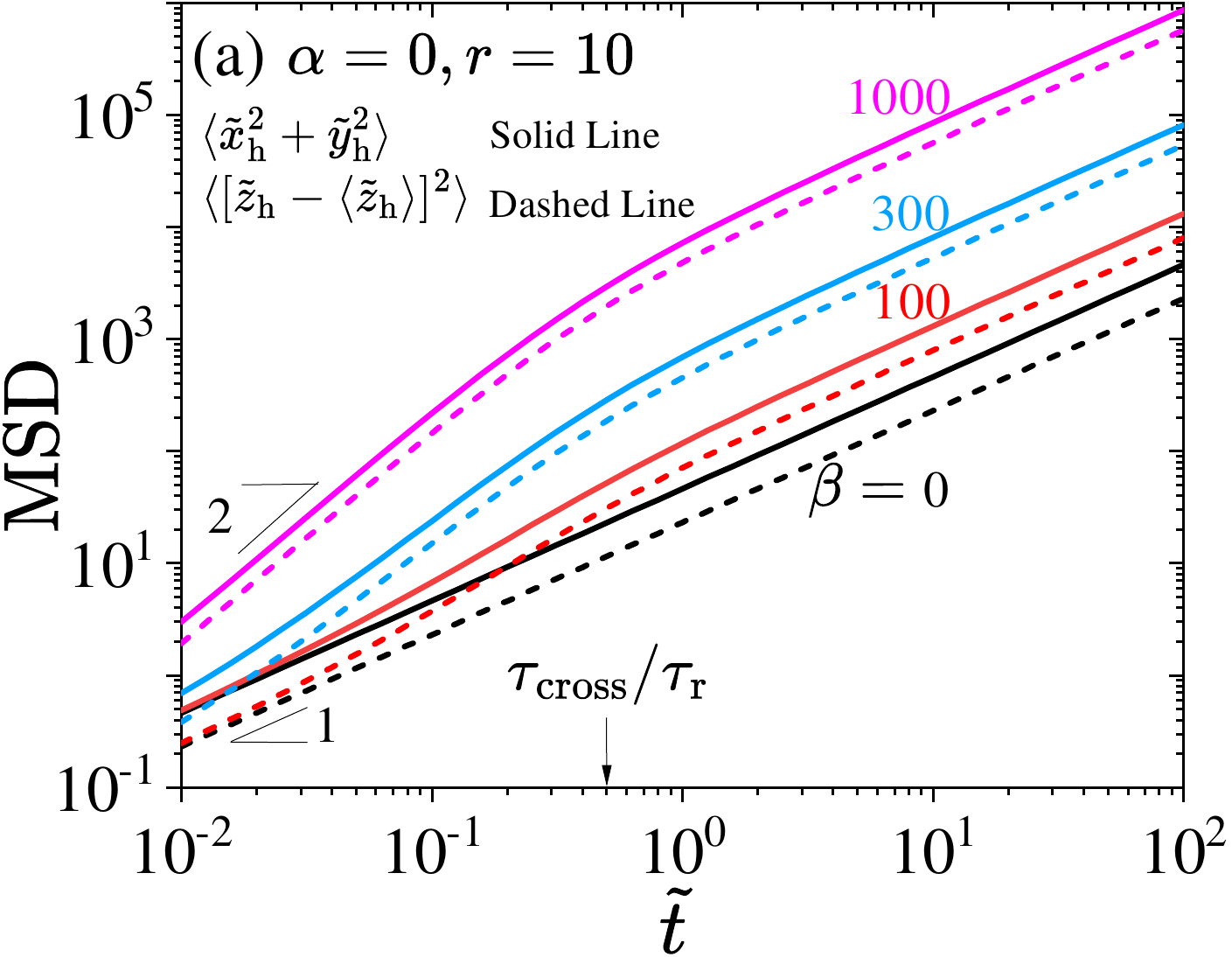}
  \includegraphics[width=0.49\textwidth]{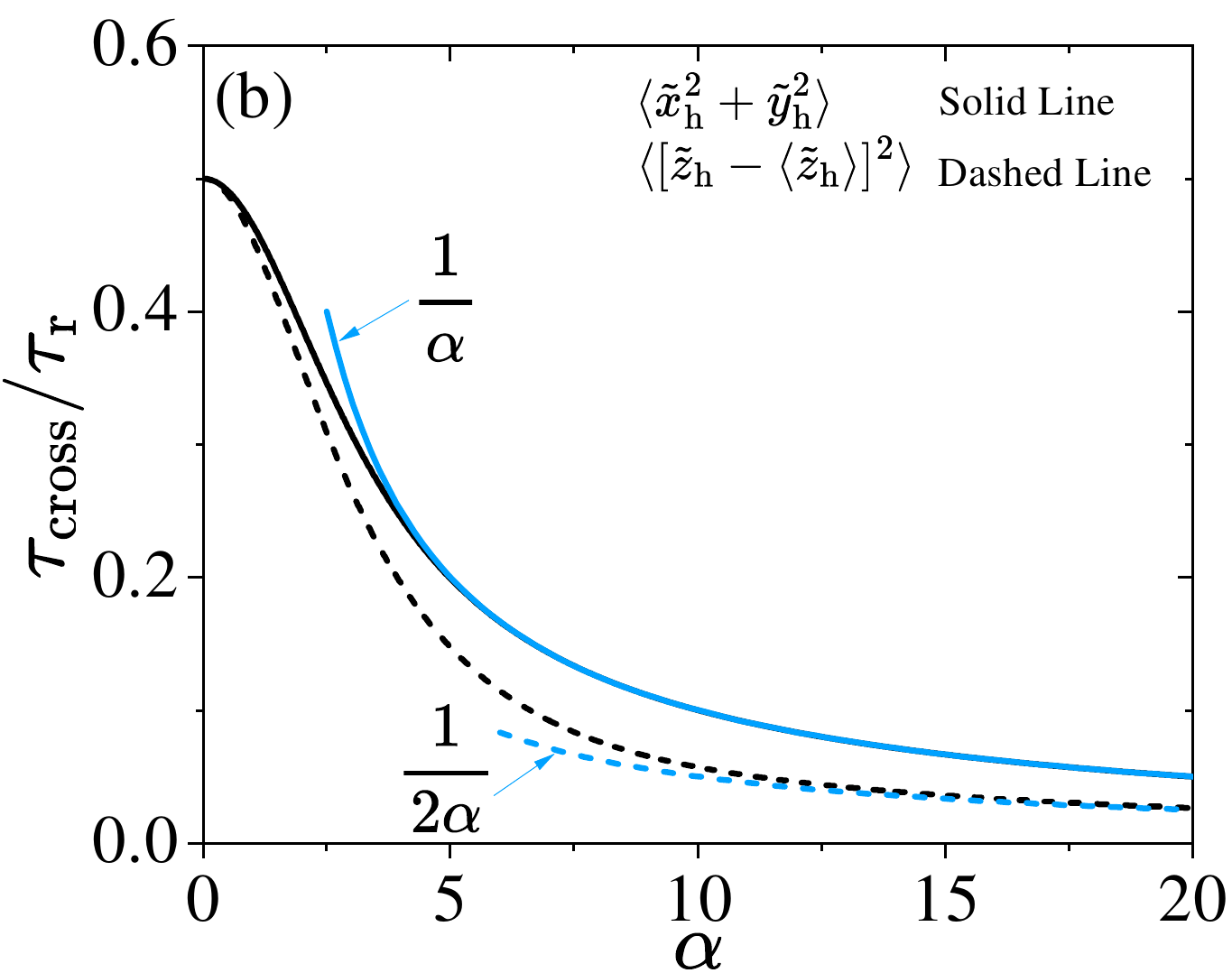}
  \caption{(a) 
 The time evolution of  
  the horizontal (solid line) and vertical (dashed line) MSD for varying sedimentation Péclet number $\beta$ ($\alpha=0$, $r=10$). The crossover time $\tau_\text{cross}$ is defined by inverse of the smallest non-zero eigenvalue of the eigenvalue equation \eqref{eignM}. 
  (b) The scaled crossover time (by $\tau_{\mathrm{r}}$) depends on $\alpha$ for both the horizontal (solid line) and vertical (dashed line) MSD.
  The analytical results for large $\alpha$ (from Eq.~\eqref{large_alpha}) are represented by blue lines. Note that $1/\alpha=\tau_\mathrm{o}/\tau_\mathrm{r}$ (see Eq.~\eqref{dimensionless_parameter}).}
\label{fig:MSDxy}
\end{figure}

\Cref{fig:MSDxy}\,(a) shows the MSD of a centrosymmetric ($\alpha=0$) spheroid of $r=10$, calculated by Eq.~\eqref{hMSD_exact_all} and the formula for the friction constants in \cref{resistance}.
For $\beta=0$, the MSD increases linearly with $t$, showing the usual diffusion behavior.
As $\beta$ increases (e.g., $100$), the MSD starts to show distinct transient behavior: the MSD first increases linearly with $\tilde{t}$, then increases in proportion to $\tilde{t}^2$, and finally shows the diffusion behavior. 
Such behavior can be understood from Eq.~\eqref{hMSD_exact}. 
For small $\tilde{t}$, the right-hand side of Eq.~\eqref{hMSD_exact} can be expanded with respect to $\tilde{t}$:
\begin{equation}
    \langle \tilde{x}_{\mathrm{h}}^2(\tilde{t})+\tilde{y}_{\mathrm{h}}^2(\tilde{t}) \rangle 
        = 4\tilde{D}^\perp\left(1+\frac{\chi}{3}\right)\tilde{t}
    + \frac{2(\beta\chi)^2}{15}\tilde{t}^2 + o(\tilde{t}^2).
\end{equation}
This gives the linear dependence for $\tilde{t} < \beta^{-2} $, and the quadratic dependence for $\tilde{t} > \beta^{-2}$. 
Therefore, the MSD crosses over from the usual diffusion scaling ($\sim \tilde{t}$) to the ballistic scaling ($\sim \tilde{t}^2$) around $\tilde{t}= \beta^{-2}$. 
For large time $\tilde{t}$, the right-hand side of Eq.~\eqref{hMSD_exact} exhibits diffusive behavior driven by both thermal diffusion and Taylor dispersion, i.e., $4\left[\tilde{D}^\perp\left(1+\chi/3\right)
    +\left(\beta \chi\right)^2/90\right] \tilde{t}$. 
There is a crossover time $\tau_\text{cross}$ from the ballistic behavior to the final diffusion behavior.
The dimensionless crossover time $\tau_\text{cross}$ (scaled by $\tau_{\mathrm{r}}$) is defined by the inverse of the smallest non-zero eigenvalue of the eigenvalue equation \eqref{eignM} with $m=1$ for horizontal dispersion and $m=0$ for vertical dispersion.

\Cref{fig:MSDxy}\,(b) shows the dimensionless crossover time as a function of $\alpha$. The crossover time indicates the dynamic transition controlled by the orientation of particles, while the orientation dynamics is affected by both rotational diffusion and gravitational torque. 
For large $\alpha$, the rotation is governed by gravitational torque, and the crossover time $\tau_\mathrm{cross}$ approaches the reorientation time $\tau_\mathrm{o}$. As can be observed in \cref{fig:MSDxy}\,(b), $\tau_\text{cross}/\tau_{\mathrm{r}}$ approaches $1/\alpha$ and $1/(2\alpha)$ for the horizontal and vertical MSD, respectively. Equation~\eqref{large_alpha} clearly indicates the relevant time scales ($1/\alpha$ and $1/(2\alpha)$) for large $\alpha$ (note that $1/\alpha=\tau_\mathrm{o}/\tau_\mathrm{r}$; see Eq.~\eqref{dimensionless_parameter}). In the absence of Brownian motion, $\tau_\mathrm{o}$ diverges as $\alpha \to 0$, which explains why the settling behavior of non-Brownian particles is sensitive to centre offsets.
If Brownian motion is present, however, for small $\alpha$ the rotation is governed by rotational diffusion and the crossover time $\tau_\mathrm{cross}$ tends to $\tau_\mathrm{r}/2$. The divergence is smoothed out by rotational diffusion, which means that the settling behavior of Brownian particles is no longer sensitive to centre offsets.

As $\beta$ increases to infinity, we can recover the non-Brownian limit. To analyze the situation, we define  $\tilde{t}_{\mathrm{s}} = t / \tau_{\mathrm{s}}$  (i.e., $\tilde{t}_{\mathrm{s}} = \mathrm{\beta} \, \tilde{t}$).
Equations~\eqref{Rxy} and \eqref{Fmt} in the limit $\beta \to \infty$ are
\begin{equation}
    \frac{\partial \left\langle \tilde{x}_{\mathrm{h}}^2+\tilde{y}_{\mathrm{h}}^2 \right\rangle}{\partial \tilde{t}_{\mathrm{s}}} 
     = 2\chi
    \mathcal{H}^1, \qquad \frac{\partial \mathcal{H}^1}{\partial \tilde{t}_{\mathrm{s}}} = \frac{2\chi}{15}, \label{limRxy}
\end{equation}
where the initial condition with the isotropic orientation distribution has been used.
Solving Eqs.~\eqref{limRxy} yields
\begin{equation}
    \lim_{\beta \to \infty} \left\langle \tilde{x}_{\mathrm{h}}^2 + \tilde{y}_{\mathrm{h}}^2 \right\rangle = \frac{2\chi^2}{15} \tilde{t}_{\mathrm{s}}^2.
\end{equation}
Both short-time and long-time diffusion behaviors in \cref{fig:MSDxy}\,(a) vanish.
The result demonstrates $\left\langle \tilde{x}_{\mathrm{h}}^2+\tilde{y}_{\mathrm{h}}^2 \right\rangle \sim \left\langle \tilde{z}_{\mathrm{h}}^2 \right\rangle \sim \tilde{t}^2$, and is valid even for a non-zero value of $\alpha$. 
This occurs because rotational Brownian motion is too weak to disrupt the gliding motion of the particle, yielding ballistic $\tilde{t}^2$ scaling.

\section{Conclusion and Discussion}

We studied the sedimentation of a Brownian axisymmetric particle with a centre offset by solving the Smoluchowski equation.
When the hydrodynamic centre and the force centre coincide ($l_\mathrm{c}=0$), we recover the Taylor dispersion of \cite{brenner1979taylor}.
When the force centre deviates from the hydrodynamic centre ($l_\mathrm{c} \neq 0$), the symmetry axis of the particle preferentially aligns with the direction of gravity.
Therefore, the sedimentation velocity increases for prolate but decreases for oblate because the friction constant $\zeta_\mathrm{t}^\parallel$ is smaller than $\zeta_\mathrm{t}^\perp$ for prolate but is larger than that for oblate.
Brenner has shown that Taylor dispersion follows a $\beta_0^2$ scaling for torque-free Brownian particles, where $\beta_0$ is the sedimentation Péclet number.
However, a significant deviation from the $\beta_0^2$ scaling appears even for a small centre offset.
Taylor dispersion is suppressed when the gravitational torque dominates over the rotational Brownian fluctuations. For sufficiently large $\beta_0$, the horizontal diffusion coefficients $D_{\text{$x$-$y$}}$ approach distinct constant values depending on the offset, while the vertical diffusion coefficients $D_z$ all converge to the same value.
In terms of gravitational torque, $D_{\text{$x$-$y$}}$ and $D_z$ asymptotically converge to $D^\perp$ and $D^\perp(1+\chi)$ as the gravitational torque goes to infinity, respectively, corresponding to a rigid particle in a gravity-aligned configuration. At intermediate gravitational torque, both the horizontal and vertical diffusion coefficients exhibit maxima.
Therefore, the Taylor dispersion could be further enhanced when a centre offset exists.
Additionally, we present a first-order perturbation analysis of the transient behavior of MSD with respect to $\alpha^2$.
When the sedimentation Péclet number goes to infinity, ballistic behavior $\left\langle \tilde{x}_{\mathrm{h}}^2 + \tilde{y}_{\mathrm{h}}^2 \right\rangle \sim \tilde{t}^2$ appears.
The crossover time reflects a dynamical transition, where rotation is governed by gravitational torque when it is large, and by rotational diffusion when the torque is small.
The ballistic behavior in MSD remains to be valid even when gravitational torque is present, indicating that the torque does not sufficiently alter sustained ballistic motion.

The most important feature of particles with centre offset is that they exhibit a larger sedimentation velocity (\cref{fig:Vz}\,(a)) than spheres and a larger diffusivity (\cref{fig:DD}\,(c) and (d)) than those with no centre offset.
The maximum diffusivity occurs around the same torque value of $\alpha \sim 2$ for both prolate and oblate spheroids, as well as for both horizontal and vertical diffusivities.
The value $\alpha \sim 2$ could be frequently accessible in practice by changing the mass distribution inside the particles. Plankton, erythrocytes, micrometre-scale sediments, etc., could use this feature to accelerate their mobility (translational velocity or diffusivity) by tuning the centre offset.

\backsection[Acknowledgements]{The authors are grateful for the support of the National Natural Science Foundation of China (NSFC), Wenzhou Institute, University of Chinese Academy of Sciences and Oujiang Laboratory.}

\backsection[Funding]{National Natural Science Foundation of China (NSFC; nos.\,22403021, 12247174, 12174390, and 12150610463) and Wenzhou Institute, University of Chinese Academy of Sciences (nos.\,WIUCASQD2022004 and WIUCASQD2020002), and Oujiang Laboratory (no.\,OJQDSP2022018).}

\backsection[Declaration of interests]{The authors report no conflict of interest.}

\appendix
\section{Onsager's variational principle}
\label{principle}

The probability distribution $\psi(\Rhc, \bm{n}, t)$ satisfies the conservation equation \citep{Doi_1986},
\begin{equation} \label{conservation}
\frac{\partial \psi}{\partial t} = 
-\frac{\partial }{\partial \Rhc}\cdot\left(\bm{u}\psi\right)
-\mathcal{D}_{\bm{n}}\cdot\left(\bm{\omega }\psi\right),
\end{equation}
and the normalization condition $\int d\Omega\psi=1$, where operator $\mathcal{D}_{\bm{n}}=\bm{n}\times\partial/\partial\bm{n}$ and $d\Omega=d\Rhc d\bm{n}$ is the volume element in the configuration space $\Omega$. 

The dissipation function of the system is the work done per unit time by the dissipative force \eqref{hydrodynamics}.
Since translation-rotation decoupling occurs at the hydrodynamic centre, the dissipation function can be expressed using the configuration distribution function $\psi$ as follows:
\begin{equation}
	\Phi=\frac{1}{2}\int d\Omega\psi\left(\bm{u}\cdot\bm{\zeta}_\mathrm{t}\cdot\bm{u}+\bm{\omega}\cdot\bm{\zeta}_\mathrm{r}\cdot\bm{\omega}\right).
\end{equation}

The free energy of the system is $A = \int d\Omega\psi\left(U+\kB T\ln\psi\right)$, where $U$ is given by Eq.~\eqref{eqn;2.3} and \eqref{center_offset} in terms of $\Rhc$ and $\bm{n}$. 
Combining the conservation equation \eqref{conservation} with integration by parts over the configuration space yields the free-energy change rate (i.e., the time derivative of the free-energy):
\begin{align}
    \dot{A} &= \int d\Omega\dot{\psi}\left[-(M-M_{\mathrm{b}})\bm{g}\cdot\Rhc -(M-M_{\mathrm{b}})l_{\mathrm{c}}\bm{g}\cdot\bm{n}+\kB T\ln\psi+\kB T\right]
    \notag \\
    &=\int d\Omega\psi\left[\bm{u}\cdot \left\{-(M-M_{\mathrm{b}})\bm{g}+\kB T\frac{\partial \ln\psi}{\partial \Rhc }\right\}
    + 
    \bm{\omega }\cdot
    \big\{-(M-M_{\mathrm{b}})l_{\mathrm{c}} \bm{n}\times\bm{g}+\kB T\mathcal{D}_{\bm{n}}\ln\psi
    \big\} 
    \right].
\end{align}

Therefore, the velocities $\bm{u}$ and $\bm{\omega}$ are obtained by minimizing the Rayleighian of the system $\mathcal{R}=\Phi+\dot{A}$, i.e.,
\begin{subequations}
\begin{align}
    \frac{\delta \mathcal{R}}{\delta \bm{u}} &= \left\{\bm{\zeta}_\mathrm{t}\cdot\bm{u} - (M-M_{\mathrm{b}})\bm{g}+\kB T\frac{\partial \ln\psi}{\partial \Rhc }\right\}\psi = 0, \\
    \frac{\delta \mathcal{R}}{\delta \bm{\omega}} &= 
    \big\{
    \bm{\zeta}_\mathrm{r}\cdot\bm{\omega} - (M-M_{\mathrm{b}})l_{\mathrm{c}} \bm{n}\times\bm{g}+\kB T\mathcal{D}_{\bm{n}}\ln\psi
    \big\}
    \psi = 0.
\end{align}
\end{subequations}

The equations can be solved using the Sherman-Morrison formula \citep{1950_Sherman} to obtain
\begin{subequations}\label{solutions}
\begin{align} 
    \bm{u} &= -\frac{1}{\zeta_\mathrm{t}^\perp}\left(\bm{\delta}+\frac{\zeta_\mathrm{t}^\perp-\zeta_\mathrm{t}^\parallel}{\zeta_\mathrm{t}^\parallel}\bm{n}\bm{n}\right) \cdot 
    \left\{ -(M-M_{\mathrm{b}})\bm{g}+\kB T\frac{\partial \ln\psi}{\partial \Rhc }\right\}, \label{solutions1}\\
    \bm{\omega} &= -\frac{1}{\zeta_\mathrm{r}^\perp} 
    \big\{-(M-M_{\mathrm{b}})l_{\mathrm{c}} \bm{n}\times\bm{g}+\kB T\mathcal{D}_{\bm{n}}\ln\psi \big\}, \label{solutions2}
\end{align}
\end{subequations}
where the contribution from the rotational component $\zeta_\mathrm{r}^\parallel$ vanishes automatically due to the vector identity $\bm{n} \times \bm{n} = \bm{0}$.
Therefore, combining Eqs.~\eqref{conservation} and \eqref{solutions} yields the Smoluchowski equation \eqref{sm_eq}.

\section{The moments of $\bm{n}$}\label{appA}

Because solving for the moment $\langle\bm{n}\bm{n}\rangle$ in Eq.~\eqref{nn} involves the moments $\langle\bm{n}\rangle$ and $\langle\bm{n}\bm{n}\bm{n}\rangle$, the same procedure used in Eq.~\eqref{evlution_eq} can be applied to derive evolution equations for correlations $\langle\bm{n}\rangle$, $\langle\bm{n}\bm{n}\bm{n}\rangle$ and $\langle\bm{n}\bm{n}\bm{n}\bm{n}\rangle$. 
We have
\begin{subequations}\label{n_s}
\begin{align} 
    \frac{\partial \langle n_i\rangle}{\partial \tilde{t}} &= \langle\tilde{\mathcal{L}}^\dagger n_i\rangle = -2\langle n_i\rangle+\alpha \left(\hat{g}_i-\hat{g}_j\langle n_jn_i\rangle\right), \label{n} \\
    \frac{\partial \langle n_in_jn_k\rangle}{\partial \tilde{t}} &= \langle\tilde{\mathcal{L}}^\dagger n_in_jn_k\rangle \notag \\
    &= -2\left(6\langle n_in_jn_k\rangle-\delta_{ij}\langle n_k\rangle-\delta_{ik}\langle n_j\rangle-\delta_{jk}\langle n_i\rangle\right) \notag \\
    &\quad
    +\alpha \left(\hat{g}_i\langle n_jn_k\rangle+\hat{g}_j\langle n_in_k\rangle+\hat{g}_k\langle n_in_j\rangle-3\hat{g}_l\langle n_ln_in_jn_k\rangle\right), \\
    \frac{\partial \langle n_in_jn_kn_l\rangle}{\partial \tilde{t}}
    &= \langle\tilde{\mathcal{L}}^\dagger n_in_jn_kn_l\rangle \notag\\
    &= -2\left(10\langle n_in_jn_kn_l\rangle-\delta_{ij}\langle n_kn_l\rangle-\delta_{ik}\langle n_jn_l\rangle-\delta_{il}\langle n_jn_k\rangle-\delta_{jk}\langle n_in_l\rangle \right.\notag \\
    &\quad
    -
    \left.
    \delta_{jl}\langle n_in_k\rangle-\delta_{kl}\langle n_in_j\rangle\right) +\alpha \left(\hat{g}_i\langle n_jn_kn_l\rangle+\hat{g}_j\langle n_in_kn_l\rangle
    \right.
    \notag \\
    &\quad
    +
    \left.
    \hat{g}_k\langle n_in_jn_l\rangle+\hat{g}_l\langle n_in_jn_k\rangle-4\hat{g}_m\langle n_mn_in_jn_kn_l\rangle\right), \label{nnnn}
\end{align}
\end{subequations}
where $i,j,k,l,m \in \{x,y,z \}$.

We solve the equations \eqref{n_s} and \eqref{nn} subject to initial conditions using the Laplace transform $\mathcal{T}[f(\tilde{t})]=\int_0^\infty d\tilde{t} e^{-\varsigma \tilde{t}}f(\tilde{t})$. 
The transformed equations are
\begin{subequations}
\begin{align}
    \varsigma T_i &= -2T_i+\alpha \left(\frac{1}{\varsigma}\hat{g}_i-T_{ij}\hat{g}_j\right), \label{T1}\\
    \varsigma T_{ij} &= \frac{1}{3}\delta_{ij}-2\left(3T_{ij}-\frac{1}{\varsigma}\delta_{ij}\right)+\alpha \left(T_i\hat{g}_j+T_j\hat{g}_i-2T_{ijk}\hat{g}_k\right), \label{T2}\\
    \varsigma T_{ijk} &= -2\left(6T_{ijk}-\delta_{ij}T_k-\delta_{ik}T_j-\delta_{jk}T_i\right) \notag \\
    &\quad+\alpha \left(T_{ij}\hat{g}_k+T_{ik}\hat{g}_j+T_{jk}\hat{g}_i-3T_{ijkl}\hat{g}_l\right), \label{T3}\\
    \varsigma T_{ijkl} &= \frac{1}{15}\left(\delta_{ij}\delta_{kl}+\delta_{il}\delta_{jk}+\delta_{ik}\delta_{jl}\right)-2\left(10T_{ijkl}-\delta_{ij}T_{kl}-\delta_{ik}T_{jl}-\delta_{il}T_{jk}-\delta_{jk}T_{il}\right.\notag\\
    &\quad
    - \left. \delta_{jl}T_{ik}-\delta_{kl}T_{ij}\right)+\alpha \left(T_{ijk}\hat{g}_l+T_{ijl}\hat{g}_k+T_{ikl}\hat{g}_j+T_{jkl}\hat{g}_i-4\hat{g}_mT_{ijklm}\right),\label{T4}
\end{align}
\end{subequations}
where $T_i = \mathcal{T}[\langle n_i\rangle]$ (at least order of $\alpha^1$), $T_{ij} = \mathcal{T}[\langle n_in_j\rangle]$ (at least order of $\alpha^0$), $T_{ijk} = \mathcal{T}[\langle n_in_jn_k\rangle]$ (at least order of $\alpha^1$), and $T_{ijkl} = \mathcal{T}[\langle n_in_jn_kn_l\rangle]$ (at least order of $\alpha^0$). 
However, solving for any given moment invariably involves higher-order moments.
Here, we retain only terms up to $\alpha^2$ in Eq.~\eqref{T2}. 
This implies we retain $T_i$ and $T_{ijk}$ to first-order in $\alpha$, while keeping $T_{ij}$ and $T_{ijkl}$ to zeroth-order in $\alpha$.
The zeroth-order expressions for $T_{ij}$ and $T_{ijkl}$ are given by:
\begin{subequations}\label{Teven_all}
\begin{align}\label{Teven}
    T_{ij} &= \frac{1}{3\varsigma}\delta_{ij} +o(\alpha^0), \\
    T_{ijkl} &= \frac{1}{15\varsigma}\left(\delta_{ij}\delta_{kl}+\delta_{ik}\delta_{jl}+\delta_{il}\delta_{jk}\right) +o(\alpha^0).\label{Teven2}
\end{align}
\end{subequations}
The first-order expressions for $T_i$ and $T_{ijk}$, derived by inserting Eq.~\eqref{Teven_all} into Eqs.~\eqref{T1} and \eqref{T3}, are given as follows:
\begin{subequations}\label{Todd_all}
\begin{align}
     T_i &= \frac{2\alpha}{3\varsigma(\varsigma+2)}\hat{g}_i +o(\alpha^1), \label{Todd1}\\
     T_{ijk} &= \frac{2\alpha }{15\varsigma(\varsigma+2)}\left(\delta_{ij}\hat{g}_k+\delta_{ik}\hat{g}_j+\delta_{jk}\hat{g}_i\right) +o(\alpha^1). \label{Todd}
\end{align}
\end{subequations}

Therefore, the second-order expression for $T_{ij}$, derived by inserting Eq.~\eqref{Todd_all} into Eq.~\eqref{T2}, is given by:
\begin{equation} \label{Tnn}
     T_{ij} = \frac{1}{3\varsigma}\delta_{ij}+\frac{4\alpha^2}{5\varsigma(\varsigma+2)(\varsigma+6)}\left(\hat{g}_i\hat{g}_j-\frac{1}{3}\delta_{ij}\right)+o(\alpha^2).
\end{equation}
The inverse Laplace transform gives the solution~\eqref{nn_sol}.

\section{The moments of $\bm{n}$ and $\tRhc$ }\label{appC}

Applying Eq.~\eqref{evlution_eq} to Eq.~\eqref{RRxy} and to the square of Eq.~\eqref{Rz} yields
\begin{subequations}
\begin{align}
    \frac{\partial \left\langle \tilde{x}_{\mathrm{h}}^2+\tilde{y}_{\mathrm{h}}^2 \right\rangle}{\partial \tilde{t}} 
    &= 
  \left\langle\tilde{\mathcal{L}}^\dagger\left\{(\bm{\delta}-\hat{\bm{g}}\hat{\bm{g}})\cdot[\tRhc(\tilde{t})-\tRhc(0)] \right\}^2 \right\rangle 
    \notag \\
    &= 
    2\tilde{D}^\perp\left[2+\chi
    \big(1- \hat{\bm{g}}\cdot\left\langle \bm{n}\bm{n} \right\rangle\cdot\hat{\bm{g}} \big)\right]
    +2\beta  \chi
    \mathcal{H}^1,  \label{Rxy}\\
    \frac{\partial \langle \left[\tilde{z}_{\mathrm{h}}-\langle \tilde{z}_{\mathrm{h}}\rangle
\right]^2\rangle
    }{\partial \tilde{t}} 
    &= 
    \left\langle     \tilde{\mathcal{L}}^\dagger\left\{\hat{\bm{g}}\cdot[\tRhc(\tilde{t})-\tRhc(0)]-\hat{\bm{g}}\cdot\left\langle\tRhc(\tilde{t})-\tRhc(0)\right\rangle\right\}^2
    \right\rangle \notag\\
    &= 
    2\tilde{D}^\perp \big(  1+\chi\hat{\bm{g}}\cdot\left\langle \bm{n}\bm{n} \right\rangle\cdot\hat{\bm{g}}
    \big)+2\beta\chi\mathcal{V}^1,  \label{Rzz}
\end{align}
\end{subequations}
where $\mathcal{H}^1=\left\langle\hat{\bm{g}}\cdot\bm{n}\bm{n}
\cdot(\bm{\delta}-\hat{\bm{g}}\hat{\bm{g}})
\cdot[\tRhc(\tilde{t})-\tRhc(0)]\right\rangle$, $\mathcal{V}^1=\left\langle\hat{\bm{g}}\cdot\bm{n}\bm{n}
\cdot\hat{\bm{g}}\hat{\bm{g}}
\cdot[\tRhc(\tilde{t})-\tRhc(0)]\right\rangle-\hat{\bm{g}}\cdot\left\langle\bm{n}\bm{n}\right\rangle
\cdot\hat{\bm{g}}\hat{\bm{g}}
\cdot\left\langle\tRhc(\tilde{t})-\tRhc(0)\right\rangle$.

Here, we focus on calculating the correlation $\mathcal{H}^1$ (and $\mathcal{V}^1$) using two equivalent approaches.
One approach is an integral method that uses the Green's function to provide an approximate solution valid across the entire range of $\alpha$.
The other approach is a differential method that employs an iterative procedure to yield exact solutions up to a finite order in $\alpha$.
\\
(1) Eigenfunction expansion method: differentiating $\mathcal{H}^1$ (and $\mathcal{V}^1$) with respect to the past time $\tilde{t}'$ yields
\begin{subequations}
\begin{align}
    \frac{\partial 
     }{\partial \tilde{t}'}\big\langle \hat{\bm{g}}\cdot\bm{n}\bm{n}\cdot(\bm{\delta}-\hat{\bm{g}}\hat{\bm{g}})\cdot(\tRhc-\tRhc')\big\rangle&=\left\langle\tilde{\mathcal{L}}^{\dagger\prime}\hat{\bm{g}}\cdot\bm{n}\bm{n}\cdot(\bm{\delta}-\hat{\bm{g}}\hat{\bm{g}})\cdot(\tRhc-\tRhc')\right\rangle \notag\\
    &=-\beta \chi\left\langle \hat{\bm{g}}\cdot\bm{n}\bm{n}\cdot(\bm{\delta}-\hat{\bm{g}}\hat{\bm{g}})\cdot\bm{n}'\bm{n}'\cdot\hat{\bm{g}}\right\rangle, \label{F1_eq}\\
    \frac{\partial 
    }{\partial \tilde{t}'}\big\langle \hat{\bm{g}}\cdot\bm{n}\bm{n}\cdot\hat{\bm{g}}\hat{\bm{g}}\cdot(\tRhc-\tRhc')\big\rangle &=\left\langle\tilde{\mathcal{L}}^{\dagger\prime}\hat{\bm{g}}\cdot\bm{n}\bm{n}\cdot\hat{\bm{g}}\hat{\bm{g}}\cdot(\tRhc-\tRhc')\right\rangle \notag\\
    &=-\beta\hat{\bm{g}}\cdot\left\langle\bm{n}\bm{n}\right\rangle\cdot\hat{\bm{g}} -\beta \chi\left\langle \hat{\bm{g}}\cdot\bm{n}\bm{n}\cdot\hat{\bm{g}}\hat{\bm{g}}\cdot\bm{n}'\bm{n}'\cdot\hat{\bm{g}}\right\rangle, \label{V1_eq}
\end{align}
\end{subequations}
where $\tilde{\mathcal{L}}^{\dagger\prime}$ denotes the operator $\tilde{\mathcal{L}}^\dagger$ acting on functions defined over the configuration $\tilde{\Omega}'$. 
Integrating both sides of Eqs.~\eqref{F1_eq} and \eqref{V1_eq} yields the solutions $\mathcal{H}^1  = 2\beta\chi\Xi(\tilde{t};\alpha)$ and $\mathcal{V}^1  = \beta\chi\Theta(\tilde{t};\alpha)$, respectively, with
\begin{subequations}
\begin{gather}
	\Xi(\tilde{t};\alpha) =
		\frac{1}{2}
		\int_0^{\tilde{t}}d\tilde{t}'\left\langle \hat{\bm{g}}\cdot\bm{n}\bm{n}\cdot(\bm{\delta}-\hat{\bm{g}}\hat{\bm{g}})\cdot\bm{n}'\bm{n}'
		\cdot\hat{\bm{g}}\right\rangle, \label{F1_eq_sol}\\
 	\Theta(\tilde{t};\alpha) =
		\int_0^{\tilde{t}}d\tilde{t}'\left[\left\langle \hat{\bm{g}}\cdot\bm{n}\bm{n}\cdot\hat{\bm{g}}\hat{\bm{g}}\cdot\bm{n}'\bm{n}'
		\cdot\hat{\bm{g}}\right\rangle-\hat{\bm{g}}\cdot\left\langle\bm{n}\bm{n}\right\rangle\cdot\hat{\bm{g}}\hat{\bm{g}}\cdot\left\langle \bm{n}'\bm{n}' \right\rangle\cdot\hat{\bm{g}}\right], \label{F1z_eq_sol}
\end{gather}
\end{subequations}
where the solution of \eqref{DRz} has been used in \eqref{F1z_eq_sol}.
This solution demonstrates that the particle displacement correlation can be obtained by time-integrating the translational velocity within the Langevin framework, where the velocity's drift term scales proportionally with the translational resistance coefficients.
Since translational resistance coefficients depend on particle orientation, the net displacement accumulates over the particle's entire orientation history.
Combining the solution in Eq.~\eqref{F1_eq_sol} with Eq.~\eqref{Rxy} yields \eqref{Rxy_sol}.
Similarly, combining the solution in Eq.~\eqref{F1z_eq_sol} with Eq.~\eqref{Rzz} yields \eqref{Rz_sol}.

In particular, the orientation-history integral $\Xi(\tilde{t};\alpha)$ (or $\Theta(\tilde{t};\alpha)$) converges to a constant $\Xi_{\mathrm{ss}}(\alpha)$ (or $\Theta_\mathrm{ss}(\alpha)$) at long times for fixed $\alpha$, where $\Xi_{\mathrm{ss}}(\alpha)$ and $\Theta_{\mathrm{ss}}(\alpha)$ are defined by Eqs.~\eqref{eigenequation} and \eqref{eigenequation2}.
Using the Green's function from Eq.~\eqref{Green_n} and incorporating the spherical coordinate relationships $n_x=\sin\theta\cos\phi$ and $n_z=\cos\theta$, we derive
\begin{subequations}\label{eigenequation_sol}
\begin{align}
    \Xi_\mathrm{ss}(\alpha) &=\sum_{p=1}^\infty \left(\int_0^{2\pi} d\phi\int_0^\pi d\theta\sin\theta \psi_p(\theta,\phi)\sin\theta\cos\phi\cos\theta\right)^2 \frac{1}{\lambda_p} \label{Xi_sol}, \\
    \Theta_\mathrm{ss}(\alpha)
    &=\sum_{p=1}^\infty \left(\int_0^{2\pi} d\phi\int_0^\pi d\theta\sin\theta \psi_p(\theta,\phi)\cos^2\theta\right)^2 \frac{1}{\lambda_p} \label{Theta_sol}.
\end{align}
\end{subequations}
Note that the $p=0$ term vanishes because $\psi_0$ is independent of $\phi$ in Eq.~\eqref{Xi_sol} and cancellation appears in Eq.~\eqref{Theta_sol}.
For $p \ne 0$ ($p=1,2,\cdots,g$), we assumed that each eigenfunction $\psi_p(\theta,\phi)$ is approximated by a linear combination of spherical harmonics $Y_l^m(\theta,\phi)$ with weight function $\psi_\mathrm{ss}^{1/2}(\theta)$, truncated to $g$ terms, i.e.,
\begin{subequations}
\begin{equation}
	\label{eignPsi}
    \psi_p(\theta,\phi) = 
    \psi_\mathrm{ss}^{1/2}(\theta)
    \sum_{l=0}^g\sum_{m=-l}^l a_{l,m}^pY_l^m(\theta,\phi),
\end{equation}
with
\begin{equation}
    Y_l^m(\theta,\phi)=\sqrt{\frac{(2 l+1) (l-m)!}{4 \pi  (l+m)!}}
    P_l^m(\cos \theta ) e^{\mathrm{i} m\phi},
\end{equation}
\end{subequations}
where $P_l^m(\cos \theta)$ are the associated Legendre functions. 
A technical consideration is that the operator $\tilde{\mathcal{L}}_\mathrm{sp}$ in the eigenequation \eqref{eigen_equation} is non-Hermitian.
To achieve a stable numerical algorithm, the operator can be transformed into a Hermitian operator by introducing the weight function $\psi_\mathrm{ss}^{1/2}(\theta)$ in Eq.~\eqref{eignPsi} \citep{risken1989fokker}.
Additionally, because $n_x$ depends on $\cos\phi$ in Eq.~\eqref{Xi_sol}, only the $m=1$ term survives in the summation over $m$ due to the orthogonality of trigonometric functions.
However, because $n_z$ is independent of $\phi$ in Eq.~\eqref{Theta_sol}, only the $m=0$ term survives in the summation over $m$ for the vertical dispersion.
Therefore, by inserting the eigenfunction form from Eq.~\eqref{eignPsi}, with $m=1$ for horizontal dispersion and $m=0$ for vertical dispersion, into the eigenequation \eqref{eigen_equation}, we obtain the first $g$ approximate solutions $\lambda_p^m$ and coefficients $a_{i,m}^p$ from
\begin{align} \label{eignM}
 \sum_{j=1}^g \mathcal{M}_{ij}^ma_{j,m}^p
 =\lambda_p^m a_{i,m}^p, \qquad (m=0,1).
\end{align}
The eigenvectors $a_{i,m}^p$ are orthonormalised, i.e., $\sum_{i=1}^ga_{i,m}^pa_{i,m}^q=\delta_{p,q}$. 
Here $\mathcal{M}_{ij}^m$ is the symmetric part ($(\bm{M}+\bm{M}^T)/2$) of the following $g \times g$ matrix:
\begin{subequations}
\begin{align}
	M_{kl}^0&= \left[k(k+1)+\frac{(k^2+k-1)\alpha^2}{2(2k-1)(2k+3)}\right] \delta_{k,l}+2\alpha\sqrt{\frac{(k+1)^2}{(2k+1)(2k+3)}}\delta_{k,l-1} \notag\\
	&\quad-\frac{\alpha^2}{2}\sqrt{\frac{(k+1)^2 (k+2)^2}{(2 k+1) (2 k+3)^2 (2 k+5)}}\delta_{k,l-2}, \qquad (k,l = 0,1,2,\dots,g-1),\\
	M_{kl}^1&= \left[k(k+1)+\frac{(k^2+k)\alpha^2}{2(2k-1)(2k+3)}\right] \delta_{k,l}+2\alpha\sqrt{\frac{k(k+2)}{(2k+1)(2k+3)}}\delta_{k,l-1}\notag\\
	&\quad-\frac{\alpha^2}{2}\sqrt{\frac{k (k+1) (k+2) (k+3)}{(2 k+1) (2 k+3)^2 (2 k+5)}}\delta_{k,l-2}, \qquad (k,l = 1,2,3,\dots,g),
\end{align}
\end{subequations}
where $\delta_{k,l}$ is the Kronecker delta. In this work, $g=35$ is sufficient to obtain numerically converged solutions. 

\noindent
(2) Iterative method: the equivalent differential equation for $\mathcal{H}^1$ can be derived by directly applying Eq.~\eqref{evlution_eq} to the correlation function $\mathcal{H}^m=\langle(\hat{\bm{g}}\cdot\bm{n})^m\bm{n}\cdot(\bm{\delta}-\hat{\bm{g}}\hat{\bm{g}})\cdot[\tRhc(\tilde{t})-\tRhc(0)]\rangle$:
\begin{align}\label{Fmt}
    \frac{\partial \mathcal{H}^m}{\partial \tilde{t}} &= (m-1)m\mathcal{H}^{m-2}-(m+1)(m+2)\mathcal{H}^m+\alpha \left[m\mathcal{H}^{m-1}-(m+1)\mathcal{H}^{m+1}\right]\notag\\
    &\quad+\beta  \chi\left(\langle(\hat{\bm{g}}\cdot\bm{n})^{m+1}\rangle-\langle(\hat{\bm{g}}\cdot\bm{n})^{m+3}\rangle\right),
\end{align}
where $m=0,1,2,\dotsc$. 
Therefore, the moments $\langle(\hat{\bm{g}}\cdot\bm{n})^m\rangle$ govern translational diffusion.
Similarly, we define $\mathcal{F}^m=\langle(\hat{\bm{g}}\cdot\bm{n})^m\hat{\bm{g}}\cdot[\tRhc(\tilde{t})-\tRhc(0)]\rangle$. 
Thus, $\mathcal{V}^1 = \mathcal{F}^2-\langle(\hat{\bm{g}}\cdot\bm{n})^2\rangle\mathcal{F}^0$. 
Applying Eq.~\eqref{evlution_eq} to $\mathcal{F}^m$ yields
\begin{align}\label{Fm_v}
    \frac{\partial \mathcal{F}^m}{\partial \tilde{t}} &= m\left[(m-1)\mathcal{F}^{m-2}-(m+1)\mathcal{F}^m+\alpha (\mathcal{F}^{m-1}-\mathcal{F}^{m+1})\right]\notag\\
    &\quad+\beta \left[\langle(\hat{\bm{g}}\cdot\bm{n})^m\rangle+\chi\langle(\hat{\bm{g}}\cdot\bm{n})^{m+2}\rangle\right],
\end{align}
where $m=0,1,2,\dotsc$.

In the long-time limit, we expect $\partial\mathcal{H}^m/\partial\tilde{t}=0$. 
Setting $m=0$ in Eq.~\eqref{Fmt} then yields an equation for $\mathcal{H}^1$.
Therefore, $\Xi_\mathrm{ss}(\alpha)$ admits the alternative expression:
\begin{align}\label{F1s}
    \Xi_\mathrm{ss}(\alpha) &= \frac{\langle(\hat{\bm{g}}\cdot\bm{n})^1\rangle_\mathrm{ss}-\langle(\hat{\bm{g}}\cdot\bm{n})^3\rangle_\mathrm{ss}}{2\alpha}-\frac{f(\alpha)}{\sinh\alpha},
\end{align}
where
\begin{subequations}
\begin{gather}
    \langle(\hat{\bm{g}}\cdot\bm{n})^m\rangle_\mathrm{ss} = \lim_{\tilde{t} \to \infty} \langle(\hat{\bm{g}}\cdot\bm{n})^m\rangle = \frac{\alpha}{2\sinh\alpha}  \int_{-1}^{1} d\xi e^{\alpha\xi}\xi^{m}, \label{gn_m}\\
    f(\alpha) = \frac{\sinh\alpha}{\beta \chi\alpha}\mathcal{H}^0_\mathrm{ss} =  \sum_{i=1,\text{odd}}^\infty p_0^i \alpha^i ,
\end{gather}
\end{subequations}
with coefficients determined iteratively by
\begin{subequations}
\begin{align}
    p_m^i &= \frac{1}{(m+1)(m+2)}\left[(m-1)mp_{m-2}^i+mp_{m-1}^{i-1}-(m+1)p_{m+1}^{i-1}+c_m^i\right], \\
    p_m^0 &= \frac{1}{3(m+2)(m+4)}, \\
    c_m^i &= \frac{2}{i!(i+m+2)(i+m+4)}.
\end{align}
\end{subequations}
The two approaches in Eqs.~\eqref{eigenequation_sol} and \eqref{F1s} yield equivalent results if $g \to \infty$.

Modeling the transient behavior of $\mathcal{H}^m$ requires the time evolution of $\langle(\hat{\bm{g}}\cdot\bm{n})^m\rangle$ in Eq.~\eqref{Fmt}. 
Again, solving for the current moment of order $m$ invariably involves the higher-order moment $m+1$. 
Therefore, applying the same procedure used in Eqs.~\eqref{Teven}--\eqref{Todd}, the solution for $\langle(\hat{\bm{g}}\cdot\bm{n})^m\rangle$ up to order of $\alpha^2$ can be obtained either from Eq.~\eqref{n_s} via contractions with $\hat{\bm{g}}$, or directly from Eq.~\eqref{evlution_eq}.
The initial conditions are set to the equilibrium values corresponding to zero gravity, i.e., the equilibrium moments are $\langle(\hat{\bm{g}}\cdot\bm{n})^m\rangle_\mathrm{eq} = 1/(m+1)$ for even $m$ and $0$ for odd $m$, which can be determined from $\left.\langle(\hat{\bm{g}}\cdot\bm{n})^m\rangle_\mathrm{ss}\right|_{\alpha=0}$. 
Using Laplace transforms, the time evolutions of $\langle(\hat{\bm{g}}\cdot\bm{n})^m\rangle$ up to the order of $\alpha^2$ are given by:
\begin{equation} \label{gn}  \mathcal{T}\left[\langle(\hat{\bm{g}}\cdot\bm{n})^m\rangle\right] = 
  \begin{cases}
      C_m^0(\varsigma)+C_m^2(\varsigma)\alpha^2, & m=0,2,4,\dotsc \\[2pt]
      C_m^1(\varsigma)\alpha , &
      m=1,3,5,\dotsc
  \end{cases},
\end{equation}
where
\begin{subequations}
\begin{align}
	C_m^0 &= \frac{1}{(m+1)\varsigma}, \label{Cm0}\\
	C_m^1 &= \frac{2}{(m+2)\varsigma(\varsigma+2)}, \\
	C_m^2 &= \frac{4m}{(m+1)(m+3)\varsigma(\varsigma+2)(\varsigma+6)}. \label{Cm2}
\end{align}
\end{subequations}

Therefore, the transient behavior of $\mathcal{H}^m$ can be calculated using the solution in Eq.~\eqref{gn}.
The solution for $\mathcal{H}^m$ up to order of $\alpha^2$ is obtained by
\begin{subequations}
\begin{equation}\label{TFm}
  \mathcal{T}\left[\mathcal{H}^m\right] =
  \begin{cases}
      \beta  \chi P_m^1(\varsigma)\alpha , & m=0,2,4, \dotsc \\[2pt]
      \beta \chi \left[P_m^0(\varsigma)+P_m^2(\varsigma)\alpha^2\right], & m=1,3,5, \dotsc
   \end{cases}
    .
\end{equation}
Applying a similar procedure to $\mathcal{F}^m$, we obtain
\begin{equation} \label{TVm}
  \mathcal{T}\left[\mathcal{F}^m\right]
  = 
  \begin{cases}
    \beta \left[Q_m^0(\varsigma,\chi)+Q_m^2(\varsigma,\chi)\alpha^2\right], & m=0,2,4,\cdots \\[2pt]
    \beta Q_m^1(\varsigma,\chi)\alpha , & m=1,3,5,\cdots
    \end{cases}
    ,
\end{equation}
\end{subequations}
where the coefficients are determined iteratively by
\begin{subequations}
\begin{align}
	P_m^0 &= \frac{1}{\varsigma+(m+1)(m+2)}\left[(m-1)mP_{m-2}^0+C_{m+1}^0-C_{m+3}^0\right], \\
	P_m^1 &= \frac{1}{\varsigma+(m+1)(m+2)}\left[(m-1)mP_{m-2}^1+C_{m+1}^1-C_{m+3}^1+mP_{m-1}^0-(m+1)P_{m+1}^0\right], \\
	P_m^2 &= \frac{1}{\varsigma+(m+1)(m+2)}\left[(m-1)mP_{m-2}^2+C_{m+1}^2-C_{m+3}^2+mP_{m-1}^1-(m+1)P_{m+1}^1\right],
\end{align}
\begin{align}
	Q_m^0 &= \frac{1}{\varsigma+m(m+1)}\left[(m-1)mQ_{m-2}^0+\left(C_m^0+\chi C_{m+2}^0\right)\right], \label{Pm0}\\
	Q_m^1 &= \frac{1}{\varsigma+m(m+1)}\left[(m-1)mQ_{m-2}^1+\left(C_m^1+\chi C_{m+2}^1\right)+m\left(Q_{m-1}^0-Q_{m+1}^0\right)\right], \\
	Q_m^2 &= \frac{1}{\varsigma+m(m+1)}\left[(m-1)mQ_{m-2}^2+\left(C_m^2+\chi C_{m+2}^2\right)+m\left(Q_{m-1}^1-Q_{m+1}^1\right)\right]. \label{Pm2}
\end{align}
\end{subequations}
Here, we explicitly present selected results from Eqs.~\eqref{TFm} and \eqref{TVm}
\begin{subequations} \label{TFH}
\begin{align} 
    \mathcal{T}\left[\mathcal{H}^0\right] &= 2\beta  \chi \frac{\varsigma+10}{15\varsigma(\varsigma+2)^2 (\varsigma+6)}\alpha ,  \\
    \mathcal{T}\left[\mathcal{H}^1\right] &= 2\beta  \chi \left[\frac{1}{15\varsigma(\varsigma+6)}+\frac{-11\varsigma^2+46\varsigma+472}{105\varsigma(\varsigma+2)^2(\varsigma+6)^2(\varsigma+12)} \alpha^2\right], \label{Fn1}
\end{align}
\begin{align}
    \mathcal{T}\left[\mathcal{F}^0\right] &= \beta \left[\frac{3+\chi }{3 \varsigma^2} +\frac{8\chi }{15 \varsigma^2 (\varsigma+2) (\varsigma+6)}\alpha^2\right], \label{F0}\\
    \mathcal{T}\left[\mathcal{F}^1\right] &= 4\beta \frac{\varsigma^2(5+2\chi )+5\varsigma(7+3\chi )+10(3+\chi )}{15 \varsigma^2 (\varsigma+2)^2(\varsigma+6)}\alpha , \\
    \mathcal{T}\left[\mathcal{F}^2\right] &= \beta \left[\frac{\varsigma(5+3\chi )+10(3+\chi )}{15\varsigma^2(\varsigma+6)}\right. \notag \\
    &\quad\left.+8\frac{\varsigma^3(21+8\chi )+4\varsigma^2(91+47\chi )+4\varsigma(357+212\chi )+336(3+2\chi )}{105\varsigma^2(\varsigma+2)^2(\varsigma+6)^2(\varsigma+12)} \alpha^2\right]. \label{F2}
\end{align}
\end{subequations}

\section{Resistance functions for the prolate and oblate spheroids} \label{resistance}

One typical application is for spheroids with the surface function
\begin{equation}
    \frac{x^2}{a^2}+\frac{y^2}{b^2}+\frac{z^2}{c^2}=1. \label{spheroids}
\end{equation}
Analytical resistance functions for prolate (where $a>b=c$, aspect ratio $r= a/c$) and oblate (where $a=b>c$, aspect ratio $r= c/a$) spheroids are tabulated in Tables 3.4 and 3.6 of \citep{Kim_1991}.
Here, we maintain a constant volume, i.e., $abc=L^3$. 
Hence, for prolate spheroids, $a=Lr^{2/3}$, while for oblate spheroids, $a=Lr^{-1/3}$. 
Therefore, the translational and rotational resistance coefficients depend solely on the aspect ratio.
For spheres ($r=1$), the resistance coefficients are $\zeta_\mathrm{t}^\parallel=\zeta_\mathrm{t}^\perp=6\pi \eta L$ and $\zeta_\mathrm{r}^\parallel=\zeta_\mathrm{r}^\perp=8\pi \eta L^3$, where $\eta$ is the viscosity of the surrounding Newtonian fluid.
For prolate spheroids ($r>1$), explicit analytical expressions are
\begin{subequations}
\begin{align}
    \zeta_\mathrm{t}^\parallel(r) &= 6\pi \eta Lr^{2/3}\frac{8}{3} \left(r^2-1\right)^{3/2}\left[r \left(2r^2-1\right) \ln \left(\frac{r+\sqrt{r^2-1}}{r-\sqrt{r^2-1}}\right)-2 r^2 \sqrt{r^2-1}\right]^{-1}, \\
    \zeta_\mathrm{t}^\perp(r) &= 6\pi \eta Lr^{2/3}\frac{16}{3} \left(r^2-1\right)^{3/2}\left[r \left(2 r^2-3\right) \ln \left(\frac{r+\sqrt{r^2-1}}{r-\sqrt{r^2-1}}\right)+2 r^2 \sqrt{r^2-1}\right]^{-1}, \\
    \zeta_\mathrm{r}^\parallel(r) &= 8\pi \eta L^3r^2\frac{4}{3} \left(r^2-1\right)^{3/2}\left[2r^4 \sqrt{r^2-1}-r^3 \ln \left(\frac{r+\sqrt{r^2-1}}{r-\sqrt{r^2-1}}\right)\right]^{-1}, \\
    \zeta_\mathrm{r}^\perp(r) &= 8\pi \eta L^3r^2\frac{4}{3} \left(r^2-1\right)^{3/2} \left(r^2+1\right)\left[ r^3 \left(2 r^2-1\right) \ln \left(\frac{r+\sqrt{r^2-1}}{r-\sqrt{r^2-1}}\right)-2 r^4 \sqrt{r^2-1}\right]^{-1}.
\end{align}
\end{subequations}
For oblate spheroids ($r<1$), explicit analytical expressions are
\begin{subequations}
\begin{align}
    \zeta_\mathrm{t}^\parallel(r) &= 6\pi \eta Lr^{-1/3}\frac{4}{3} \left(1-r^2\right)^{3/2}\left[\left(1-2 r^2\right) \mathrm{arccot}\left(\frac{r}{\sqrt{1-r^2}}\right)+r\sqrt{1-r^2}\right ]^{-1}, \\
    \zeta_\mathrm{t}^\perp(r) &= 6\pi \eta Lr^{-1/3}\frac{8}{3} \left(1-r^2\right)^{3/2}\left[\left(3-2 r^2\right) \mathrm{arccot}\left(\frac{r}{\sqrt{1-r^2}}\right)-r \sqrt{1-r^2}\right ]^{-1}, \\
    \zeta_\mathrm{r}^\parallel(r) &= 8\pi \eta L^3r^{-1}\frac{2}{3} \left(1-r^2\right)^{3/2}\left[ \mathrm{arccot}\left(\frac{r}{\sqrt{1-r^2}}\right)- r \sqrt{1-r^2}\right ]^{-1}, \\
    \zeta_\mathrm{r}^\perp(r) &= 8\pi \eta L^3r^{-1}\frac{2}{3} \left(1-r^2\right)^{3/2} \left(1+r^2\right)\left[ \left(1-2 r^2\right) \mathrm{arccot}\left(\frac{r}{\sqrt{1-r^2}}\right)+ r\sqrt{1-r^2}\right ]^{-1}.
\end{align}
\end{subequations}

\bibliographystyle{jfm}
\bibliography{main}

\begin{thebibliography}{19}
\expandafter\ifx\csname natexlab\endcsname\relax\def\natexlab#1{#1}\fi
\def\au#1{#1} \def\ed#1{#1} \def\yr#1{#1}\def\at#1{#1}\def\jt#1{\textit{#1}} \def\bt#1{#1}\def\bvol#1{\textbf{#1}} \def\vol#1{#1} \def\pg#1{#1} \def\publ#1{#1}\def\arxiv#1{#1}\def\org#1{#1}\def\st#1{\textit{#1}}

\bibitem[Angle {\em et~al.\/}(2024)Angle, Rau \& Byron]{PhysRevFluids.9.070501}
{\sc \au{Angle, Brandon~R.}, \au{Rau, Matthew~J.} \& \au{Byron, Margaret~L.}} \yr{2024}  \at{Settling of nonuniform cylinders at intermediate \uppercase{R}eynolds numbers}.  \jt{Phys. Rev. Fluids}  \bvol{9}~(7),  \pg{070501}.

\bibitem[Brenner(1979)]{brenner1979taylor}
{\sc \au{Brenner, Howard}} \yr{1979}  \at{Taylor dispersion in systems of sedimenting nonspherical \uppercase{B}rownian particles. \uppercase{I}. homogeneous, centrosymmetric, axisymmetric particles}.  \jt{J. Colloid Interf. Sci.}  \bvol{71}~(2),  \pg{189--208}.

\bibitem[Brenner(1981)]{brenner1981taylor}
{\sc \au{Brenner, Howard}} \yr{1981}  \at{Taylor dispersion in systems of sedimenting nonspherical \uppercase{B}rownian particles: \uppercase{II}. homogeneous ellipsoidal particles}.  \jt{J. Colloid Interf. Sci.}  \bvol{80}~(2),  \pg{548--588}.

\bibitem[Brenner \& Condiff(1972)]{BRENNER1972228}
{\sc \au{Brenner, Howard} \& \au{Condiff, Duane~W}} \yr{1972}  \at{Transport mechanics in systems of orientable particles. \uppercase{III}. arbitrary particles}.  \jt{J. Colloid Interf. Sci.}  \bvol{41}~(2),  \pg{228--274}.

\bibitem[Brenner \& Edwards(1993)]{brenner1993macrotransport}
{\sc \au{Brenner, Howard} \& \au{Edwards, David, A}} \yr{1993} {\em Macrotransport Processes\/}.  \publ{Elsevier}.

\bibitem[Dill \& Brenner(1983)]{dill1983general}
{\sc \au{Dill, Loren~H} \& \au{Brenner, Howard}} \yr{1983}  \at{A general theory of \uppercase{T}aylor dispersion phenomena. \uppercase{VI}. \uppercase{L}angevin methods}.  \jt{J. Colloid Interf. Sci.}  \bvol{93}~(2),  \pg{343--365}.

\bibitem[Doi \& Edwards(1986)]{Doi_1986}
{\sc \au{Doi, Masao} \& \au{Edwards, Sam~F}} \yr{1986} {\em The theory of polymer dynamics\/}.  \publ{Oxford University Press}.

\bibitem[Frankel(1991)]{frankel1991approach}
{\sc \au{Frankel, I}} \yr{1991}  \at{The approach to normality in the dispersion of sedimenting nonspherical \uppercase{B}rownian particles}.  \jt{J. Colloid Interf. Sci.}  \bvol{142}~(1),  \pg{179--203}.

\bibitem[Goren(1979)]{GOREN1979209}
{\sc \au{Goren, Simon~L}} \yr{1979}  \at{Effective diffusivity of nonspherical, sedimenting particles}.  \jt{J. Colloid Interf. Sci.}  \bvol{71}~(2),  \pg{209--215}.

\bibitem[Harvey \& Garcia de~la Torre(1980)]{harvey1980coordinate}
{\sc \au{Harvey, Steven} \& \au{Garcia de~la Torre, Jose}} \yr{1980}  \at{Coordinate systems for modeling the hydrodynamic resistance and diffusion coefficients of irregularly shaped rigid macromolecules}.  \jt{Macromolecules}  \bvol{13}~(4),  \pg{960--964}.

\bibitem[Kim \& Karrila(1991)]{Kim_1991}
{\sc \au{Kim, S.} \& \au{Karrila, S.~J.}} \yr{1991} {\em Microhydrodynamics: Principles and selected applications\/}.  \publ{Boston: Butterworth-Heinemann}.

\bibitem[Pagitsas {\em et~al.\/}(1986{\natexlab{{\em a\/}}})Pagitsas, Nadim \& Brenner]{10.1063/1.450305}
{\sc \au{Pagitsas, Michael}, \au{Nadim, Ali} \& \au{Brenner, Howard}} \yr{1986{\natexlab{{\em a\/}}}}  \at{Projection operator analysis of macrotransport processes}.  \jt{J. Chem. Phys.}  \bvol{84}~(5),  \pg{2801--2807}.

\bibitem[Pagitsas {\em et~al.\/}(1986{\natexlab{{\em b\/}}})Pagitsas, Nadim \& Brenner]{pagitsas1986multiple}
{\sc \au{Pagitsas, Michail}, \au{Nadim, A~a} \& \au{Brenner, H}} \yr{1986{\natexlab{{\em b\/}}}}  \at{Multiple time scale analysis of macrotransport processes}.  \jt{Physica. A}  \bvol{135}~(2-3),  \pg{533--550}.

\bibitem[Risken(1989)]{risken1989fokker}
{\sc \au{Risken, Hannes}} \yr{1989} {\em The Fokker-Planck Equation: Methods of Solution and Applications\/}.  \publ{Springer}.

\bibitem[Roy {\em et~al.\/}(2019)Roy, Hamati, Tierney, Koch \& Voth]{Roy_Hamati_Tierney_Koch_Voth_2019}
{\sc \au{Roy, Anubhab}, \au{Hamati, Rami~J.}, \au{Tierney, Lydia}, \au{Koch, Donald~L.} \& \au{Voth, Greg~A.}} \yr{2019}  \at{Inertial torques and a symmetry breaking orientational transition in the sedimentation of slender fibres}.  \jt{J. Fluid Mech.}  \bvol{875},  \pg{576–596}.

\bibitem[Sherman \& Morrison(1950)]{1950_Sherman}
{\sc \au{Sherman, Jack} \& \au{Morrison, Winifred~J.}} \yr{1950}  \at{Adjustment of an inverse matrix corresponding to a change in one element of a given matrix}.  \jt{Ann. Math. Statistics}  \bvol{21}~(1),  \pg{124--127}.

\bibitem[Swan \& Wang(2016)]{swan2016rapid}
{\sc \au{Swan, James~W} \& \au{Wang, Gang}} \yr{2016}  \at{Rapid calculation of hydrodynamic and transport properties in concentrated solutions of colloidal particles and macromolecules}.  \jt{Phys. Fluids}  \bvol{28}~(1),  \pg{011902}.

\bibitem[Will \& Krug(2021)]{PhysRevLett.126.174502}
{\sc \au{Will, Jelle~B.} \& \au{Krug, Dominik}} \yr{2021}  \at{Rising and sinking in resonance: Mass distribution critically affects buoyancy-driven spheres via rotational dynamics}.  \jt{Phys. Rev. Lett.}  \bvol{126},  \pg{174502}.

\bibitem[Xiong {\em et~al.\/}(2024)Xiong, Seto \& Doi]{macromol4c00532}
{\sc \au{Xiong, Zhongqiang}, \au{Seto, Ryohei} \& \au{Doi, Masao}} \yr{2024}  \at{Bending--rotation coupling in the viscoelasticity of semiflexible polymers-rigorous perturbation analysis from the rod limit}.  \jt{Macromolecules}  \bvol{57}~(11),  \pg{5289--5299}.

\end{thebibliography}

\end{document}